\documentclass{emulateapj}
\usepackage{txfonts}
\usepackage{graphicx}
\usepackage{epstopdf}
\usepackage{natbib} 
\bibpunct{(}{)}{;}{a}{}{,} 
\shorttitle{Are giant tornadoes the legs of solar prominences?}
\shortauthors{Wedemeyer et al.}
\begin{document}
%
\title{Are giant tornadoes the legs of solar prominences?}

\author{Sven Wedemeyer, 
Eamon Scullion, 
Luc Rouppe van der Voort,
Antonija Bosnjak}
\affil{Institute of Theoretical Astrophysics, University of Oslo,
  P.O. Box 1029 Blindern, N-0315 Oslo, Norway}
\email{sven.wedemeyer@astro.uio.no}

\author{Patrick Antolin}
\affil{Centre for mathematical Plasma Astrophysics, Department of Mathematics, KU Leuven, Celestijnenlaan 200B, bus 2400, B-3001 Leuven, Belgium}

\date{Received April 5th, 2013; accepted July 17th, 2013}

\begin{abstract}
Observations in the 171\,\AA\ channel of the Atmospheric Imaging Assembly 
of the space-borne Solar Dynamics Observatory show tornadoes-like features 
in the atmosphere of the Sun. 
These giant tornadoes appear as dark, elongated and apparently rotating 
structures in front of a brighter background. 
This phenomenon is thought to be produced by rotating magnetic field 
structures that extend throughout the atmosphere. 
We characterize giant tornadoes through a statistical analysis of properties
like spatial distribution, lifetimes, and sizes.  
A total number of 201~giant tornadoes are detected in a period of 25~days, 
suggesting that on average about 30~events are present across the whole Sun 
at a time close to solar maximum. 
Most tornadoes appear in groups and seem to form the legs of prominences, 
thus serving as plasma sources/sinks. 
Additional H$\alpha$ observations with the Swedish 1-m Solar Telescope 
imply that giant tornadoes rotate as a structure
although clearly exhibiting a thread-like structure. 
We observe tornado groups that grow prior to the eruption of the connected 
prominence.  
The rotation of the tornadoes may progressively twist the 
magnetic structure of the prominence until it becomes unstable and erupts. 
Finally, we investigate the potential relation of giant tornadoes to other 
phenomena, which may also be produced by rotating 
magnetic field structures. 
A comparison to cyclones, magnetic tornadoes and  
spicules implies that such events are more abundant and short-lived
the smaller they are. 
This comparison might help to construct a power law for the effective atmospheric 
heating contribution as function of spatial scale.
\end{abstract} 



\section{Introduction}
\label{sec:intro}

\begin{figure}[t!]
\centering
\includegraphics[width=6.5cm]{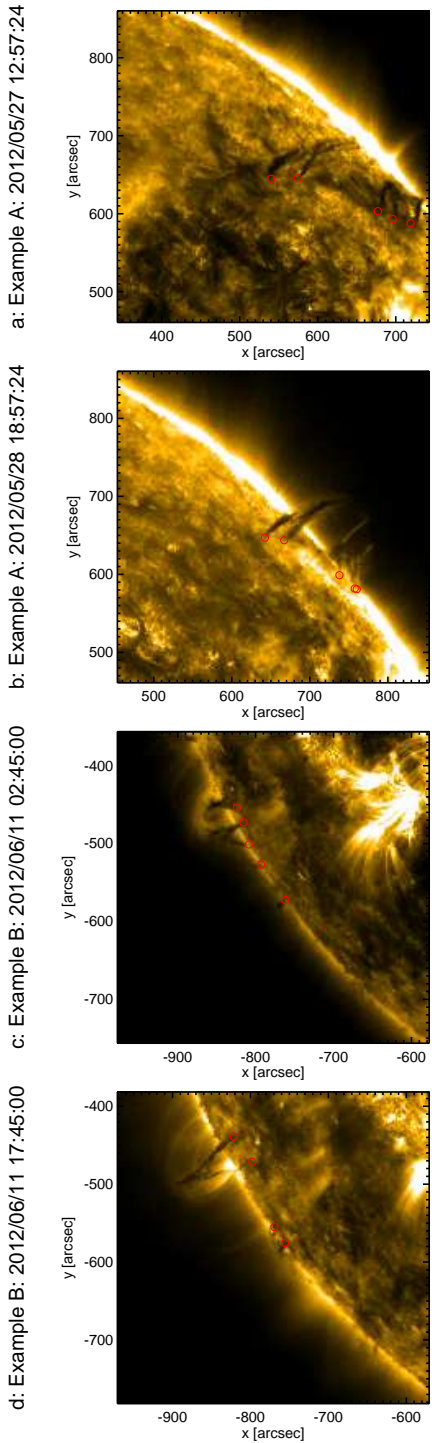}
\caption{Intensity images from the AIA channel at 171\,\AA\ showing examples of 
tornado groups. 
The group in the two upper panels consists of 5 clear tornadoes at different times, 
whereas the other group in the two lower panels has at least 4-5 tornadoes. 
The bottom panel shows the latter group shortly before eruption.
The tornado footpoints are marked by small red dots.
\label{fig:tornado}}
\end{figure}

The term `tornado' has been used repeatedly for phenomena on the Sun, in 
particular in connection with prominences \citep{1932ApJ....76....9P}, 
although the physical processes behind their formation are very different from 
those responsible for terrestrial tornadoes. 
'Tornado prominences' have been studied numerous times with 
spectroscopic observations dating back to 1868 
\citep[cf.][]{1869AN.....74..269Z, 1950PASP...62..144P}. 
Tandberg-Hansen refers to tornadoes as vertically aligned, helical structures, 
which are connected to solar prominences 
\citep[see Chap. 10 in][]{1977ASSL...69.....B}. 
The terms prominence and filament are used synonymously here since they 
describe the same phenomenon seen on-disk and at the limb, respectively
\citep[see, e.g.,][ and references therein for overview articles]{1989SoPh..119..341Z, 
1998ASPC..150...23E, 1998SoPh..182..107M, 2010SSRv..151..333M,2013SoPh..tmp..146J}. 
In the 1990s, observations of solar tornadoes, to which we refer to as `giant tornadoes' 
hereafter, have been made  by \citet{1998SoPh..182..333P} 
with the Coronal Diagnostic Spectrometer \citep[CDS,][]{1995SoPh..162..233H}
onboard the Solar and Heliospheric Observatory \citep[SOHO,][]{1995SoPh..162....1D}. 
These observations suggest that rotation may play an important role for the dynamics 
of the solar transition region  
\citep[see also][]{1997SoPh..175..457P,2000A&A...355.1152B}.  
More recently, this phenomenon was observed in detail with the 
Atmospheric Imaging Assembly (AIA) onboard the Solar Dynamics Observatory 
\citep[SDO,][]{2012SoPh..275...17L}. 
For instance, \citet{2012ApJ...752L..22L} detected plasma that moves along spiral
paths with an apparent rotation that lasted for more than three hours. 
Different types of rotation of a filament are also observed by 
\citet{2013ApJ...764...91S}. 
It is likely that giant tornadoes -- at least the majority of them -- 
are the vertical legs of prominences and filaments. 
Recently, \citet{2012ApJ...756L..41S} suggested that barbs are another observational 
imprint of giant tornadoes \citep[see also][]{2012ApJ...752L..22L}. 
Barbs are observed in the H$\alpha$ line and are known to be rooted in the
photosphere and to connect to the horizontal 'spine' of a filament in the upper atmosphere 
\citep[e.g.,][and references therein]{1998Natur.396..440Z,1998SoPh..182..107M,2013SoPh..282..147L}. 
Despite their apparent motion, it is not clear yet \citep[see, e.g.,][]{2013arXiv1307.2303P} 
if giant tornadoes truly rotate as entities although \citet{2012ApJ...761L..25O} provided 
support for this hypothesis. 
They measured Doppler shifts of $\pm 6$\,km\,s$^{-1}$ at the opposite sides 
of legs of a quiescent hedgerow prominence, which they  
interpret as rotation of the structure around an axis vertical to
the solar surface. 
\cite{2013A&A...549A.105P} suggest that tornadoes could be 
caused by the helical magnetic field of a prominence in response to the 
expansion of the corresponding cavity, whereas 
\citet{2012ApJ...756L..41S} propose that giant tornadoes can be explained 
as rotating magnetic structures driven by underlying photospheric vortex flows. 
The latter explanation matches the mechanism that has recently been found
for so-called magnetic tornadoes by \citet{2012Natur.486..505W} although 
these events, which have only been observed on-disk so far, seem to be smaller 
than the prominence-related giant tornadoes discussed here. 
Hence, it is not clear yet if the small-scale tornadoes and the giant tornadoes
are connected but the physical processes behind might be similar to some extent.  
The  observations and accompanying 3D numerical simulations by 
\citet{2012Natur.486..505W}  clearly show that magnetic tornadoes are caused 
by vortex flows in the photosphere, which force the footpoints of magnetic 
field structures to rotate. 
These events are observed with the Swedish 1-m Solar Telescope 
\citep[SST,][]{2003SPIE.4853..341S} as `chromospheric swirls'  in  image
sequences in the core of the \ion{Ca}{2}\,854.2\,nm line \citep{2009A&A...507L...9W} 
but also have an imprint in AIA/SDO images.

Here, we present a systematic analysis of giant tornadoes as they appear 
in SDO~AIA\,171\,\AA\ images regarding 
their spatial distribution and statistics of their sizes and lifetimes. 
We also compare the AIA~171\,\AA\ tornadoes to the corresponding imprints 
in other AIA~channels, magnetograms, and high-resolution SST observations.  
Based on this data, we provide support for the following hypotheses:   
(i)~Giant tornadoes are an integral part of solar prominences. 
(ii)~Giant tornadoes serve as sources and sinks of the mass flow of 
prominence material.
(iii)~They may inject helicity into the connected prominence, which can lead 
to its eruption. 
The observations are described in Sect.~\ref{sec:observ}, followed by the 
results in Sect.~\ref{sec:result}, 
discussion in Sect.~\ref{sec:disc}  and conclusions in Sect.~\ref{sec:conc}, 
respectively. 

\begin{figure}[t!]
\begin{center}
\vspace*{-5mm}
\includegraphics[width=8cm]{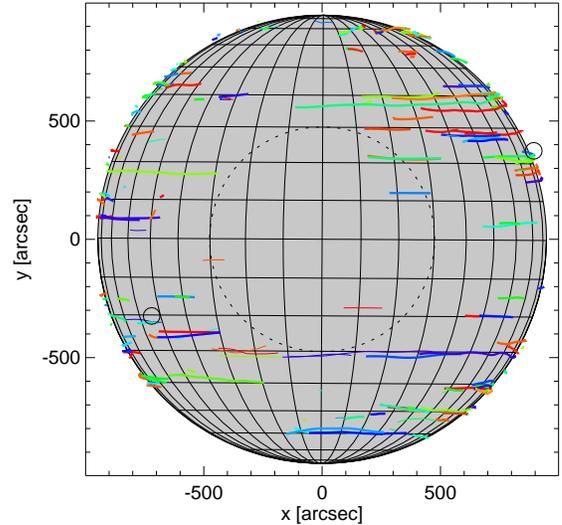}
\caption{Distribution of all detected events on the full disk of the Sun. 
The tracks of all events are plotted although not all of them exist at the same time.
The central region appears to be void due to projection effects that make it 
difficult to detect the tornado-like structure close to disk-centre.
The two open circles mark the positions of the detailed SST observations.
\label{fig:overview}}
\end{center}
\end{figure}

\begin{figure*}[t!]
\begin{center}
\includegraphics[width=15cm]{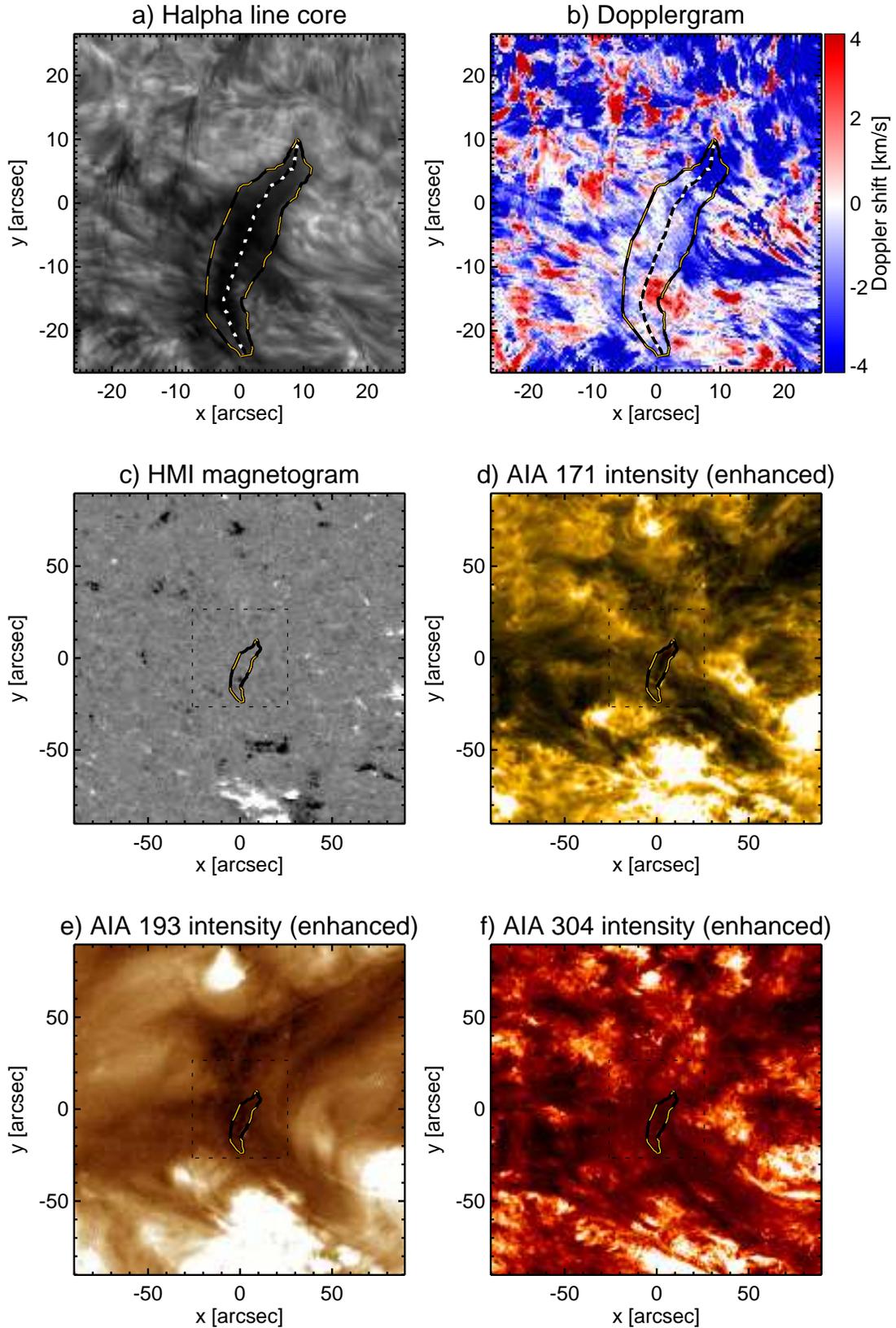}
\caption{Observation of an on-disk tornado. 
\textbf{a)}~H$\alpha$ line core image taken with SST/CRISP 
and \textbf{b)}~the corresponding Dopplergram.  
The color range of the Dopplergram is limited to $\pm 4$\,km\,s$^{-1}$, 
while the full range is $[-14.7, 11.8]$\,km\,s$^{-1}$. 
The yellow-black dashed line outline the tornado, whereas the 
black-white dotted line represents the centerline.  
The panels below display SDO images for a larger FOV: 
\textbf{c)}~HMI magnetogram, 
\textbf{d)}~AIA~171, 
\textbf{e)}~AIA~193, and 
\textbf{f)}~AIA~304. 
The dotted boxes mark the SST FOV.
\label{fig:halphadisk}}
\end{center}
\end{figure*}

\section{Observations and data reduction}
\label{sec:observ}

We analyze images taken with AIA/SDO and systematically detect `giant tornadoes' throughout 
the period from May 27$^\mathrm{th}$, 2012 to June 21$^\mathrm{st}$, 2012. 
{It is important to note that this period corresponds to a time of high solar activity.}
This period is long enough to cover roughly one solar revolution and thus results in 
a statistically significant  sample. 
The data channel at 171\,\AA\ is used for the initial detection with 
JHelioviewer\footnote{Available at http://www.jhelioviewer.org} 
and for a determination of positions and lifetimes.  
Next, the AIA images for the whole period are downloaded with a cadence of 1~hour and 
post-processed with standard SolarSoft routines to level 1.5 data. 
It includes dark current correction, flat-fielding, hot pixel correction and de-spiking, 
as well as deconvolution with a point spread function, de-rotation and scale correction.

Those events that are already present in the first time step or are still present 
in the last timestep are followed beyond the analyzed period in order to reliably 
determine their lifetime. 
The positions and the shapes of all events are evaluated throughout the time series. 
For selected events, 193\,\AA, 211\,\AA, and 304\,\AA\ images are investigated, too.  
Corresponding magnetograms from the Helioseismic and Magnetic Imager 
\citep[HMI,][]{2012SoPh..275..207S} onboard SDO 
are analysed in order to determine the photospheric magnetic field 
topology close to the tornadoes. 
Examples of tornadoes are shown in Fig.~\ref{fig:tornado}. 
They appear as dark and elongated features in front of a brighter background with 
a narrow footpoint and a more diffuse top. 
Time-series of the dark features reveal apparent rotation. 
This signature is pronounced for the majority (81.5\,\%) of all cases 
but can be more subtle for other examples, which we hereafter refer to as 
less confidently identified. 
In addition, daily full-disk \mbox{H$\alpha$} images from the Big Bear 
Solar Observatory (BBSO) are obtained from the Virtual Solar Observatory 
\citep{2004SPIE.5493..163H} for the whole observation period and searched 
for prominences.

Two of the tornadoes, which are identified in our sample, have also been 
observed with the SST on June 8th, 2012. 
One event  was observed on-disk, whereas the other  was located above the limb 
(see open circles in Fig.~\ref{fig:overview}).
The on-disk event was observed at 9:58~UT at $x=-720$\arcsec, $y=-342$\arcsec\ 
and the off-limb event at 12:24~UT at $x=894$\arcsec, $y=374$\arcsec. 
The CRISP instrument  \citep{2008ApJ...689L..69S}  was used
to scan through the H$\alpha$ line with a fine wavelength sampling of 
86\,m\AA\ (39~line positions for the on-disk event, 33 line positions for
the off-limb event). 
The seeing was very variable during the time of observation which 
inhibited extensive temporal coverage. 
Here we restrict to the analysis of the best scans. 
For the on-disk event, high-spatial resolution was achieved for the best scan
with the aid of the SST adaptive optics system 
\citep{2003SPIE.4853..370S} 
and Multi-Object Multi-Frame Blind Deconvolution image restoration
\citep{2005SoPh..228..191V}. 
The pixel scale of these frames is 0.059\arcsec. 
For more information on the optical set-up and processing, we refer 
to, e.g., \citet{2012ApJ...752..108S}. 
For the off-limb event, the adaptive optics system could not actively 
compensate for seeing and we restrict the image post-processing to the 
standard dark current and flatfield corrections.  
A pixel binning of $2\,\times\,2$ has been applied in this case.
The quality of the selected line scans was nevertheless sufficient to 
construct reliable Dopplergrams. 
The nature of the spectral line profiles, which change from 
absorption profiles on-disk to emission profiles off-limb, is taken into 
account for the determination of the Doppler shifts. 
The Doppler shifts for flat and noisy profiles, which are found in off-limb areas 
outside the tornado, are set to zero. 
Corresponding AIA images for the same time and locations
are rotated and aligned with the SST observations.
For inspection and exploration of the SST data we used the widget-based analysis
tool CRISPEX \citep{2012ApJ...750...22V}, which allows for efficient exploration of 
multi-dimensional datasets.
See Sects.~\ref{sec:sstondisk} and \ref{sec:offlimbtornado} for the results of the analysis.

\section{Analysis of tornado properties}
\label{sec:result}

\subsection{AIA and SST observations of an on-disk tornado}
\label{sec:sstondisk}

\begin{figure*}[t!]
\begin{center}
\includegraphics[width=12cm]{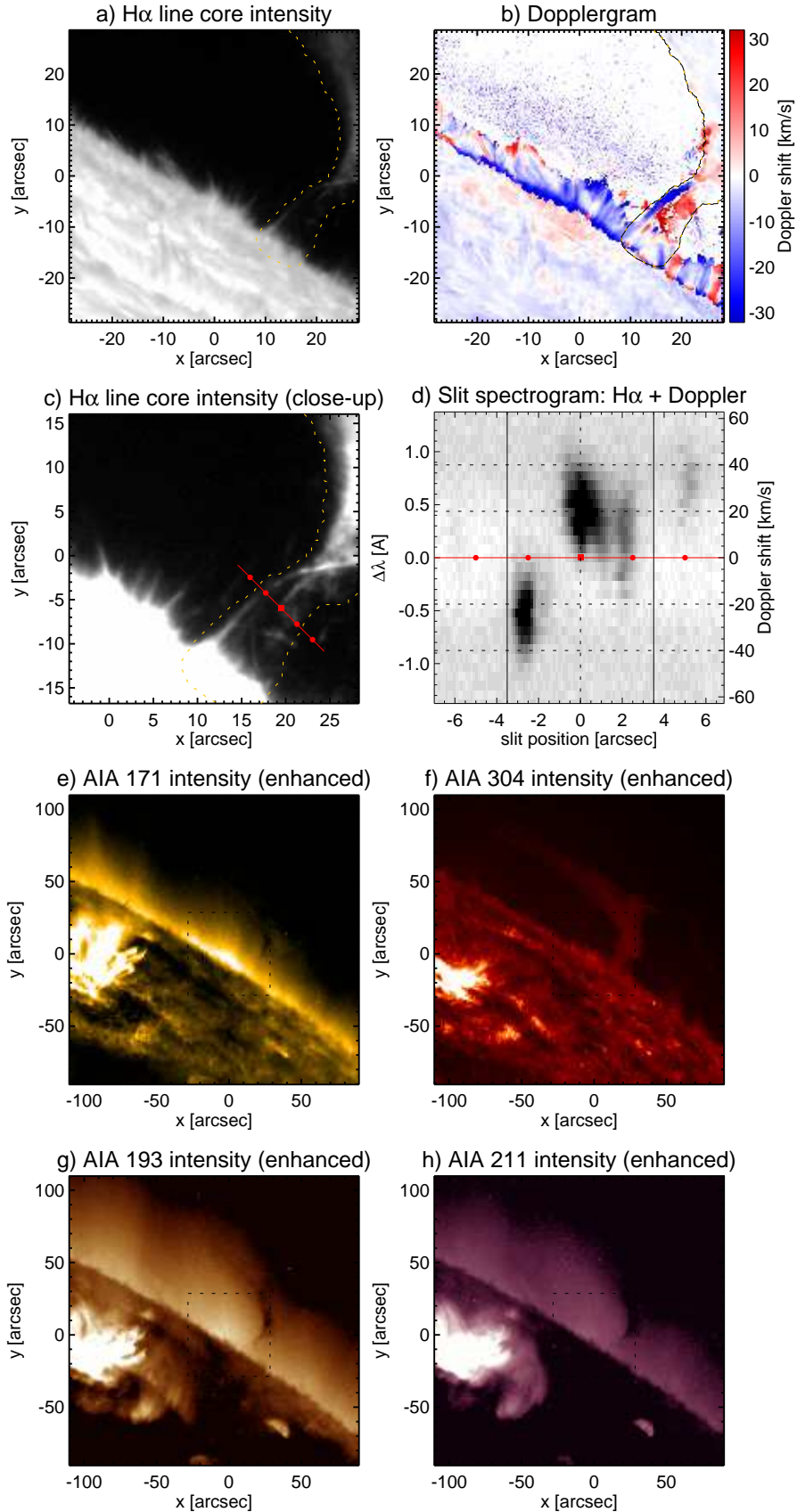}
%
\caption{Off-limb observation of a tornado. 
\textbf{a)}~H$\alpha$ line core image taken with the SST on June 8th, 2012, 
and \textbf{b)}~the corresponding Dopplergram. 
\textbf{c)}~Enhanced H$\alpha$ line core intensity for a close-up of the 
tornado base with an artificial slit (red line with markers every 
2.5\arcsec). 
\textbf{d)}~Slit spectrogram, i.e., reversed H$\alpha$ intensity as function 
of position along the slit and wavelength. 
The black vertical lines mark the tornado boundaries. 
The panels below show SDO/AIA images for a slightly larger FOV: 
\textbf{e)}~171\,\AA, 
\textbf{f)}~304\,\AA, 
\textbf{g)}~193\,\AA, 
\textbf{h)}~211\,\AA.  
The FOV from the top row is marked as dotted box. 
\label{fig:halphalimb}}
\end{center}
\end{figure*}
%
\begin{figure}[t!]
\centering
\includegraphics[width=\columnwidth]{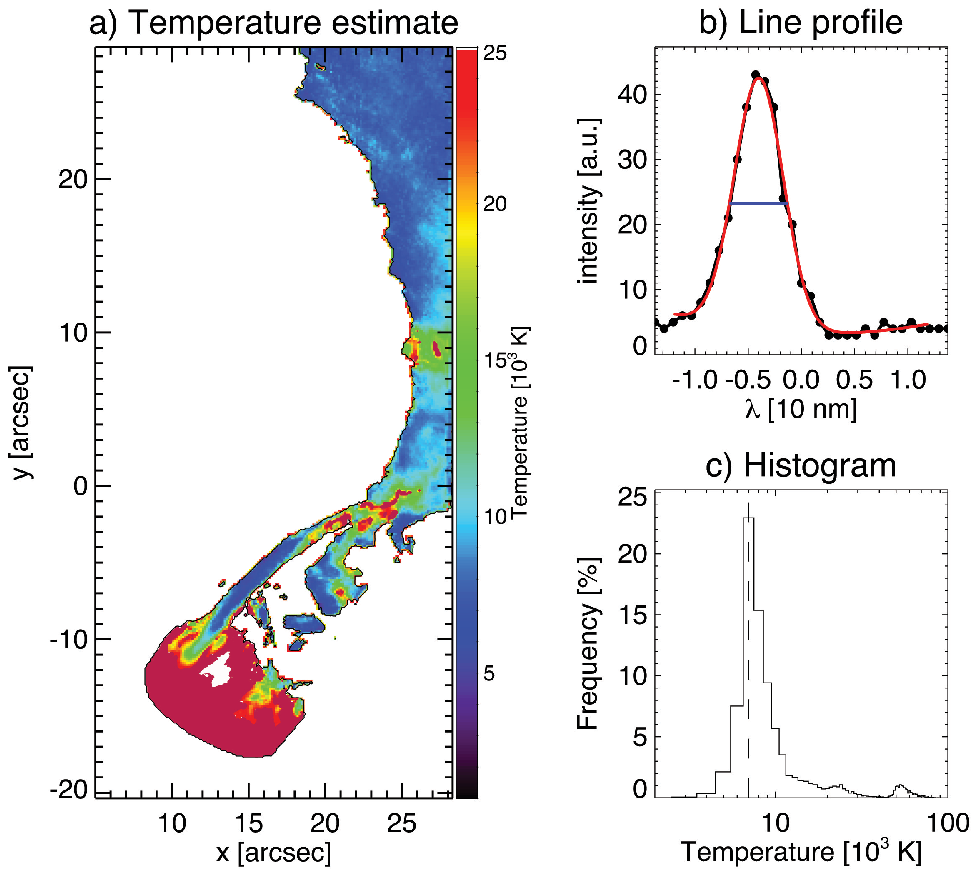}
\caption{Temperature estimate for the off-limb tornado. 
\textbf{a)}~Derived temperature in a close-up region. Pixels are white if no value 
was determined and pink if the value exceeds 20\,000\,K. 
\textbf{b)}~Intensity profile at $x = 16.1\arcsec, y = -5.3\arcsec$ (black dots) with 
Gaussian fit (red) and FWHM (horizontal blue line), 
\textbf{c)}~Histogram of the temperature values for all pixels that are not white 
in panel a. 
\label{fig:halphatemp}}
\end{figure}

The on-disk tornado is clearly visible in the H$\alpha$ line core images and also 
the AIA\,171\,\AA\ image, whereas there is only a subtle imprint in the other 
AIA channels (see Fig.~\ref{fig:halphadisk}). 
Please refer to \citet{2005ApJ...622..714A} for a detailed account on the 
formation of the intensity at 171\,\AA. 
The tornado in the H$\alpha$ image is slightly bent and has a length of 
37\arcsec\ measured along the centerline. 
The dark structure is less than 4.4\arcsec\ wide with its 
middle part having a typical width of \mbox{$\sim3.3$\arcsec}.
The imprint in the 171\,\AA\ image has a similar width, whereas the length 
is determined to only \mbox{$\sim31$\arcsec} because the thinnest parts are 
not resolved with SDO. 
The dark thin threads to the left of the tornado outline the overlying 
filament, which is visible in the 193\,\AA\ and 304\,\AA\ images.
The filament lies presumably above the tornado, which is rooted in a 
photospheric footpoint and connects to the filament threads. 
The footpoint, which we do not observe directly, is likely to be located 
close to the end of the H$\alpha$ signature 
(at $x = 9.2$\arcsec, $y = 9.7$\arcsec\ in Fig.~\ref{fig:halphadisk}a). 
The HMI magnetogram shows no clear photospheric footpoint of the tornado. 
There is only one single  magnetic point co-located with the tornado 
but it is not clear if this photospheric feature is connected to the 
chromospheric tornado signature as seen in H$\alpha$
(see Fig.~\ref{fig:halphadisk}c).

The spectral line profiles at positions inside the on-disk tornado 
have only small Doppler shifts, are mostly symmetric and have a lower 
line core intensity than the background. 
Many threads in the filament spine, which thus are not part of the tornado itself,  
can also be discerned in the Doppler map as a pattern of stripes 
almost perpendicular to the tornado axis. 
This imprint implies that the line core intensity has strong contributions from the 
filament threads, whereas the tornado only causes moderate additional 
absorption. 
We conclude that the strong contributions of the filament render 
these observations unsuitable for measuring 
velocities that could be attributed unambiguously  to the tornado.

\subsection{AIA and SST observations of an off-limb tornado}
\label{sec:offlimbtornado}
The SST observations of the off-limb tornado provide a clearer 
picture (see Fig.~\ref{fig:halphalimb}a-d). 
The corresponding AIA~171~\AA\ image  (see Fig.~\ref{fig:halphalimb}e) indeed 
shows that the SST observations coincide with a detected tornado event.
Although the lower part of the tornado is difficult to see for this 
instance in time in the AIA 171 \AA\ image, a dark absorption feature  can be 
outlined. 
There is an indication of a cavity above the tornado, like for the 
event observed by \citet{2012ApJ...752L..22L}. 
The 304\,\AA\ AIA image in Fig.~\ref{fig:halphalimb}f reveals a prominence 
bending sidewards from the tornado, roughly coinciding with the observed cavity. 
The 304\,\AA\ image further reveals a narrow base for the tornado, which extends 
$\sim$~30\arcsec above the  solar limb before it bends sidewards towards the 
pole. 
The field of view (FOV) of the SST observations includes the tornado base. 
The H$\alpha$ line core images (see Fig.~\ref{fig:halphalimb}a,c) exhibit 
thin elongated threads that extend almost vertically above the limb before
bending sidewards, too. 
These threads give the impression of a helical structure, while the lateral 
continuation resembles a 'smoke-like streamer' -- features of 'tornado 
prominences' that have been described for a long time 
\citep[e.g.][]{1950PASP...62..144P}.
The best visible thread, which is rooted at the left side of the tornado 
base, has a width of 0.4\arcsec~-~0.5\arcsec. 
This width is close to the limit of what can be resolved in this data set in 
view of the moderate seeing conditions.
The lower parts of the thread coincide with an elongated region with a 
strong blue shift in the SST Doppler map in Fig.~\ref{fig:halphalimb}b, 
although the blue-shifted region has a width of 1.6\arcsec~-~1.8\arcsec\ 
and is thus much wider than the line core thread. 
The blue shifts are on the order of $-20$\,km\,s$^{-1}$ to $-30$\,km\,s$^{-1}$. 
The remaining tornado base exhibits strong red shifts mostly between 
$10$\,km\,s$^{-1}$ and $30$\,km\,s$^{-1}$. 
Both regions extend vertically $\sim$\,12\arcsec. 
Above, mostly only weak shifts are observed. 
The Doppler signal in the lowermost part of the tornado is obscured  
by plasma in the foreground. 
The whole tornado base, which is both visible in the Dopplergram and in 
the 304\,\AA\ AIA image, has a width between 6\arcsec\ and 8\arcsec, measured 
from the outer edges of the Doppler shifted regions.  
The spectrogram in Fig.~\ref{fig:halphalimb}d) shows the H$\alpha$ 
intensity along the artificial slit, which is marked in panel~c. 
Dark shades mark the H$\alpha$ emission peak and thus allow for estimating 
the Doppler shift along the slit. 
Along the slit inside the tornado (between the vertical black lines), 
the Doppler shift changes from about  $-20$\,km\,s$^{-1}$  on one 
side to about  $+20$\,km\,s$^{-1}$ on the other side. 
At this speed, an uniformly rotating cylinder with a width of 7\arcsec\  
would revolve in $\sim 14$\,min. 
On the other hand, we expect a more complex rotation pattern given 
the fine-structure, which is clearly visible 
in the close-up of the line core image in Fig.~\ref{fig:halphalimb}c 
and also in the apparent gap in the middle of the slit in 
Fig.~\ref{fig:halphalimb}d. 
The tornado structure may rotate as  a whole but only the thin threads, 
which compose the tornado base, provide tracers of the rotation.

\label{sec:otherchannels}
The tornado structure is also visible in other AIA channels, which 
differ in the formation and the contribution to the intensity in the 
pass-bands and -- in approximation -- map plasma at different 
effective temperature. 
With the exception of the 304\,\AA\ channel in which the tornado appears 
in emission, all other AIA channels show the tornado as a dark 
absorption-like structure. 
All analysed channels show a tornado structure that is narrow at the 
base, funnels out above and continues sideways into the prominence. 
This lateral continuation is most visible in the 304\,\AA\ channel but 
appears only as a very faint streak in the 171, 193, and 211.  
The imprint in the 304\,\AA\ image has an apparent total length of 
140-150\arcsec, measured from the footpoint at the limb to the outermost 
faint end. 
This imprint has a width of $\sim$8\arcsec at the base, 14-15\arcsec\ near 
the bend-over point and 14-20\arcsec\ in the prominence part. 
The thin thread that is visible in emission in the H$\alpha$ line core image 
is also visible as a dark feature in 193\,\AA\ next to another thread that is 
better visible in 193\,\AA\  than in H$\alpha$. 
The 211\,\AA\ images are very similar. 
We measure widths for these threads, which are on the order of the pixelscale 
($\sim$0.6\arcsec) of SDO. 
The tornado structure funnels out above the base and has widths of 
5\,-\,8\arcsec\ close to the bend-over point.
This width agrees with the characteristic value of 5.7\arcsec, which 
is determined based on the 171\,\AA\ signature. 
The corresponding length is 20\arcsec.

The Doppler shifts with opposite sign at the edges of the tornado 
base strongly imply that the tornado structure rotates. 
In that picture, the plasma on one side moves towards the observer and the 
plasma on the other side of the tornado away from the observer. 
The observation that the red-shifted area is larger than the blue-shifted 
part speaks against a uniformly rotating cylinder but rather implies 
a more complex geometry like, e.g., a elliptical cross-section or a 
threaded sub-structure. 
In this regard, it should be emphasized that the observations at the limb are 
very challenging and important details may not have been resolved so far. 
The Doppler signature can nevertheless be interpreted as a rotation, 
while the amplitudes of the Doppler velocities have 
to be analysed with more caution.  
It cannot be completely ruled out that we observe two regions 
that appear close in projection but actually move away from each other at high 
speed. 
It seems however questionable that the tornado structure could remain stable 
for the observed duration in this case. 
Another possible explanation is that the Doppler signals are due to 
counter-streaming flows like they have been observed by 
\citet{1998Natur.396..440Z} in barbs and along the  spine of a prominence.  
They measure plasma speeds of $5 - 20$\,km\,s$^{-1}$, which is in the same 
range as the Doppler shifts found in this work.  
Although this possibility certainly cannot be ruled out on basis of the 
available data, the rotation of the tornado base seems to be the more likely 
explanation in view of the apparent motions in SDO image series. 
This conclusion is in line with the results by \citet{2012ApJ...761L..25O} 
for the legs of a quiescent hedgerow prominence, for which they determine 
opposite Doppler shifts of $\pm 6$\,km\,s$^{-1}$ at the opposite sides. 
This interpretation also agrees with the findings 
concerning solar tornadoes presented by \citet{2012ApJ...756L..41S}, 
who report on rotation speeds of $5 - 10$\,km\,s$^{-1}$, 
and \citet{2012Natur.486..505W}.

By calculating the width of the H$\alpha$ line profile we can further determine
upper limits for the plasma temperature inside the tornado. 
At a position of $x = 16.1\arcsec, y = -5.3\arcsec$ in the left blue-shifted 
thread, we observe a typical emission profile, which is shown in 
Fig.~\ref{fig:halphatemp}b. 
We correct the profile for straylight contributions by subtracting the intensity 
offset as determined from the outer wavelength positions and then fit the profile 
with a Gaussian. 
The line width $\Xi$ is then derived as FWHM of the Gaussian.  
The corresponding gas temperature of the emitting region in the tornado can  
be estimated according to \citet[][see their Eq.~7]{2012ApJ...745..152A} as 
\begin{equation} 
T = \frac{1}{16\,\ln 2}\ \frac{c^2\,m_\mathrm{H}}{k_\mathrm{B}}\ \left(\frac{\Xi}{\lambda_0}\right)^2\qquad,
\end{equation} 
where we ignored the microturbulence. 
Here we choose to ignore the microturbulence in order to obtain upper limits for 
the temperature. 
The resulting temperature values span a range from 4\,000\,K to 88\,000\,K, 
although the temperature is typically below 25\,000\,K in the depicted part of 
the tornado (see Fig.~\ref{fig:halphatemp}a). 
The corresponding histogram in panel~c exhibits a maximum at 7\,000\,K.
In the left blue-shifted thread, we find temperatures between 5\,000\,K and 
10\,000\,K.

\subsection{Abundance and spatial distribution.} 
\label{sec:abudis}

\begin{figure}[t!]
\begin{center}
\includegraphics[width=8cm]{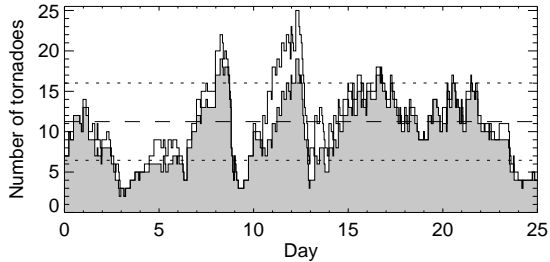}
\caption{Total number of tornadoes visible in AIA\,171\,\AA\ images as function of time. 
The grey-shades area represents the number of confidently detected tornadoes whereas 
the white areas on top also account for less confidently detected events. 
The horizontal lines represent the average (dashed) plus/minus standard deviation
(dotted) for all events. 
\label{fig:nevent}}
\end{center}
\end{figure}

A total number of 201 tornadoes is detected in the time period of 25~days. 
On average there are 11.2 events present at the same time although the total 
number varies between 2 and 25 (see Fig.~\ref{fig:nevent}).
Due to projection effects tornadoes are difficult to detect in the central parts 
of the solar disk inside a half solar radius ($R_\mathrm{c} = \frac{1}{2}\,R_\odot$, 
see dotted circle in 
Fig.~\ref{fig:overview}). 
The numbers should be multiplied by a factor to account for the 
central part and the backside of the Sun. 
We derive correction factors between $4/\sqrt{3}$ (when accounting for the observed 
area fraction on a spherical surface) and 
\slantfrac{8}{3} when using the area fraction on the projected disk instead. 
With the latter, we estimate that on average there are about 30~tornadoes present 
on the Sun at the same time during the analysed period.   
The numbers are reduced to 9.7\,events at the same time and about~26 
over the whole surface if only the confidently detected events are considered.
The variation of this number with the solar cycle has to be investigated in a 
future study.

\begin{figure}[t!]
\begin{center}
\includegraphics[width=8cm]{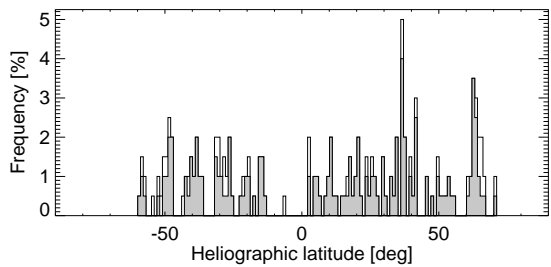}
\caption{Distribution of giant tornadoes over heliographic latitude: All events 
(white) and confident examples only (grey). 
\label{fig:spatdis}}
\end{center}
\end{figure}

The distribution of the tornado events over latitude is shown in 
Fig.~\ref{fig:spatdis}. 
They appear mostly at mid-latitudes, whereas there are essentially no giant 
tornadoes present close to the solar equator and the poles. 
This distribution resembles that of the activity belt, which suggests 
a close connection to strong magnetic field concentrations.


Based on the BBSO full-disk H$\alpha$ images, we find that  
91\,\% of all detected tornadoes are co-located with filaments or prominences. 
The remaining apparently 'isolated' tornadoes are less confident detections, except 
for two cases which are only seen shortly before disappearing at the west limb. 
It is therefore plausible to assume that all giant tornadoes are part of a filament. 
The majority of the tornadoes in our data set therefore appear in groups located along 
filament channels. 
The analysis of HMI magnetograms  reveals that these  'tornado alleys' 
are related to polarity inversion lines (see Sect.~\ref{sec:magnet}).  
The groups consist typically of three to seven coexistent tornadoes. 
Additional tornadoes can appear and disappear during the lifetime of a group, 
resulting in up to 15~tornado detections for some groups. 
We find a continuous spectrum of distances between coexistent tornadoes within 
the same group, where 83\,\% of all tornadoes are located 80\arcsec\ or less and 
71\,\% less than 50\arcsec\ from a neighbouring group member, respectively. 
In many cases, the tornado groups are co-located with `barbs' in connection 
with a filament/prominence (see Sect.~\ref{sec:disc}).

\begin{figure}[t!]
\begin{center}
\includegraphics[width=9cm]{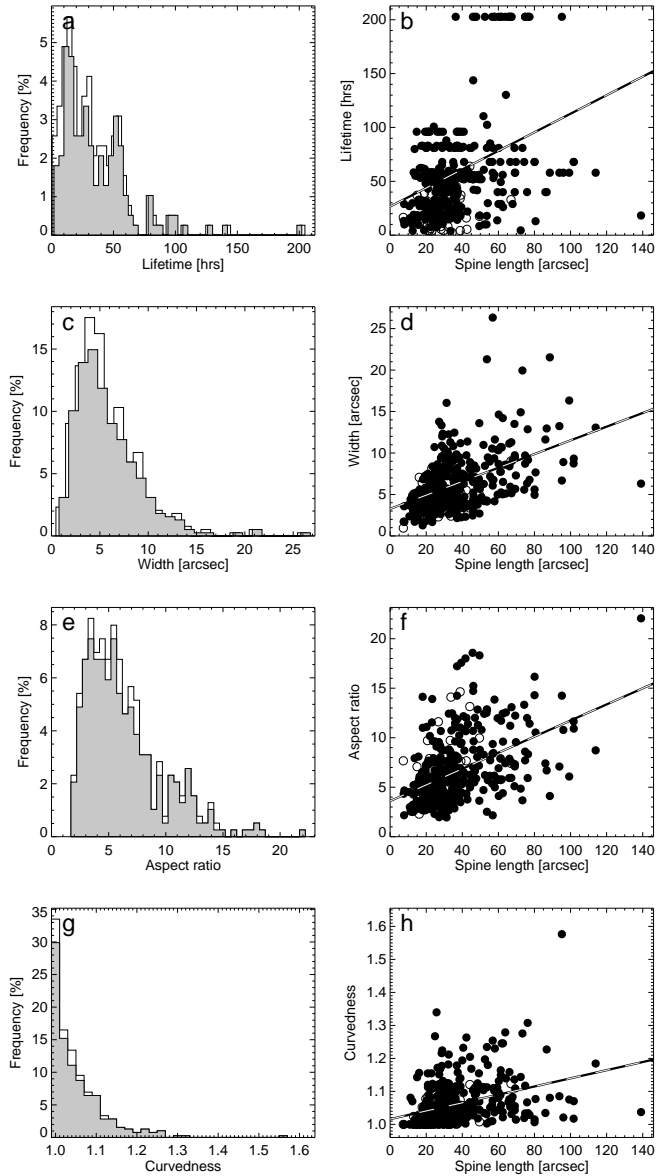}
\caption{Statistics of tornado lifetime (a,b), characteristic width (c,d), 
aspect ratio (e,f), and curvedness (g,h). 
The left column shows the histogram for all measurements (white) and 
for the confident examples (grey). 
Each property is plotted against the tornado centralline length (right column), where 
confident examples are represented by filled circles and less confident ones 
by open circles. 
A linear regression line is plotted to illustrate the trend for each property. 
\label{fig:width}}
\end{center}
\end{figure}

\subsection{Apparent tornado lifetimes in 171\,\AA\ images} 

The lifetimes of all confidently detected tornadoes range from only one hour 
to 202~hours. 
Events that existed already at the beginning of the analyzed period and 
events that lasted longer than the period have been tracked beyond the period 
and the lifetimes have been determined correspondingly. 
The histogram of the lifetimes (Fig.~\ref{fig:width}a) reveals that most 
tornadoes live for less than 60~hours. 
The average lifetime is 35\,hours with a large standard 
deviation of 27\,hours although the latter is reduced to 20~hours when excluding 
the less confidently detected examples. 
Only 15~events (7.5\,\%) last for more than 3~days and only 8~
(4\,\%) for more than 4~days, respectively. 
Only 5\,\% of all examples have been clearly visible for less than 7~hours.  
It should be emphasized that the lifetime of an observed tornado signature,
i.e., an apparently rotating plasma funnel in AIA (and SST) images, may not be  
identical to the lifetime of the rotating magnetic structure, which may produce 
the signature.  
Some tornadoes seem to appear at locations where previously another 
event had been appeared and disappeared again. 
That could be interpreted such that a magnetic `skeleton' persist for a long time, 
while it becomes only temporarily visible as a tornado. 
The connection between a tornado signature and rotating magnetic fields has been 
demonstrated by \citet{2012Natur.486..505W} for 'magnetic tornadoes', which 
in many aspects appear to be similar to the giant tornadoes discussed here. 
It is further possible that the signature gets obscured by other dynamical 
features and projection effects or is temporarily not visible in the used filter 
passband, resulting in a short apparent lifetime. 
In particular, the shortest lifetimes might thus be caused by such detection 
effects. 
Considering only a subsample of 36 events, which are all confidently detected 
and members of clear tornado groups, slightly increases the average lifetime to 
44~hours.
As will be discussed in Sect.~\ref{sec:disc}, most tornadoes seem to be connected 
to filaments, which exist for much longer than the 'lifetimes' determined here.

\begin{figure*}[t!]
\begin{center}
\includegraphics[width=12cm]{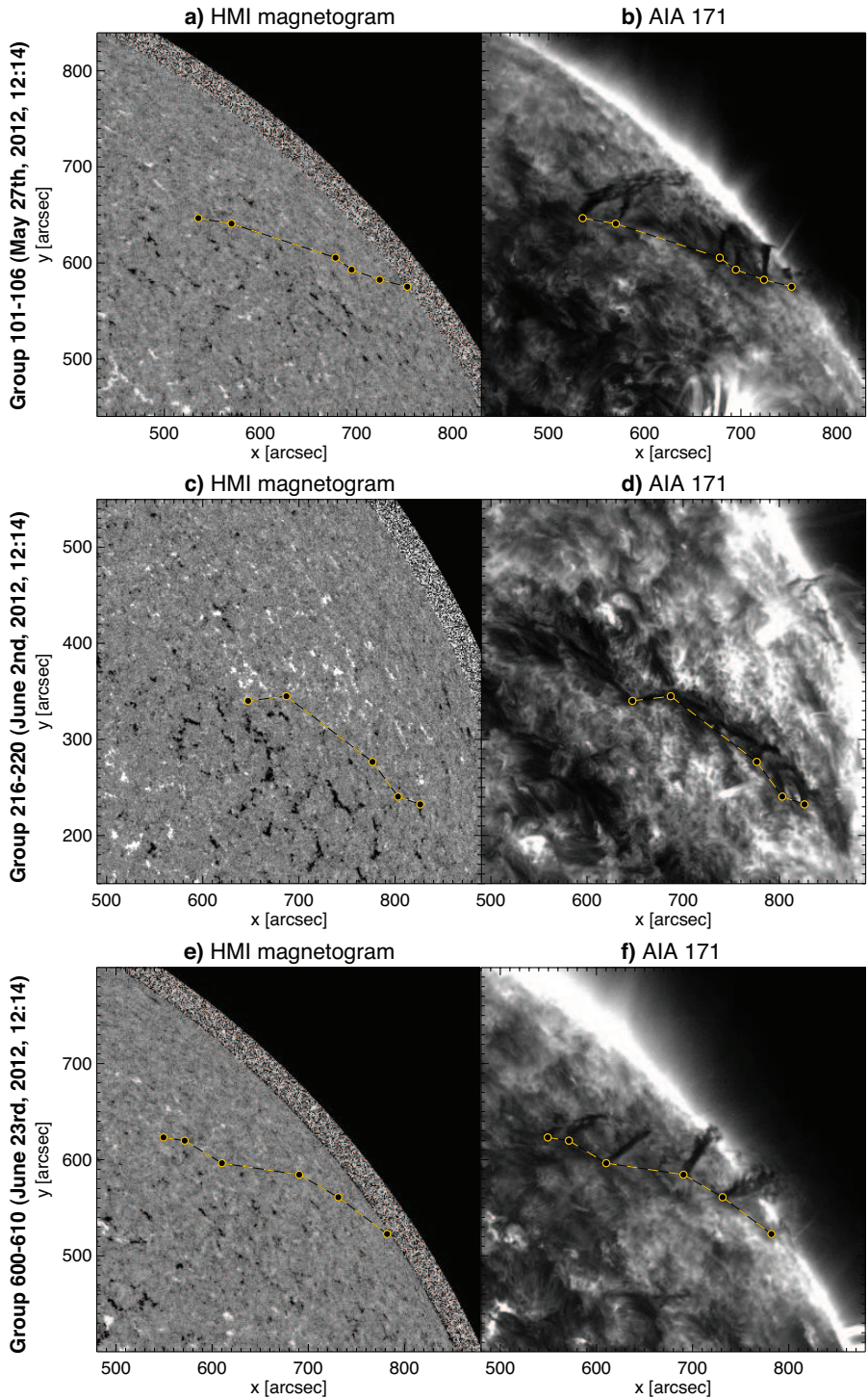}
\caption{HMI magnetograms (left column) and corresponding AIA~171 images (right 
column) for three selected tornado groups. 
The footpoints of the AIA~171 tornadoes, which are marked with circles and connected 
by lines in all panels, outline the polarity inversion lines in the magnetograms. 
\label{fig:magnetogram}}
\end{center}
\end{figure*}

\subsection{Tornado sizes in 171\,\AA\ images} 
\label{sec:tornadosize}
 
The spatial extent of the tornadoes in the 171\,\AA\ images is measured for all 
confidently detected events  
for at least one or several snapshots, representing different stages in their 
evolution. 
The resulting data set consists of 392~tornado shapes.
For each example, the footpoint and the centerline of the tornado 
followed along the structure are determined. 
Most tornadoes have centerline lengths between 10\arcsec\ and 100\arcsec\ 
($\sim7$\,Mm - $75$\,Mm). 
It is not always obvious where a tornado starts and where it ends. 
Adding the topmost part of the observed structure to the tornado instead 
of the prominence can lead to an overestimation of the lengths of 
the longest  centerlines.  
The continuation into the prominence is better seen in 304\,\AA\ images 
(see Sect.~\ref{sec:otherchannels}).
The width of the tornado is then measured along the centerline and usually 
increases from the footpoint outwards.  
With these uncertainties in mind, we determine the average tornado length to 
be $36.3\arcsec~\pm~19.0$\arcsec\ (26.3\,Mm~$\pm 13.8$\,Mm). 

We derive a characteristic width for each example by averaging the width along 
the middle part of the centerline. 
The characteristic widths extend over a large range between mostly 
2\arcsec and 16\arcsec\ (1.5\,Mm - 11.6\,Mm) 
with a few smaller and larger cases (see Fig.~\ref{fig:width}a). 
On average the width is (6.1\,$\pm$\,3.3)~\arcsec\ (4.4\,Mm~$\pm$\,2.4\,Mm).

In general, the width increases with the length, which implies that tornadoes 
scale to different sizes (see linear regression in Fig.~\ref{fig:width}d).
About 80\,\% of all shapes have an aspect ratio, i.e. centerline length divided 
by characteristic width, of 8 or less (Fig.~\ref{fig:width}e-f). 
Many examples appear to have a rather straight centerline, whereas others are  
significantly curved. 
We define the curvedness of a tornado as the ratio of the path length of the 
spine and the distance between the footpoint and the most distant point. 
The resulting histogram is shown in Fig.~\ref{fig:width}g. 
In almost all cases the curvedness stays below 1.3. 
There is also a trend of larger  curvedness with increasing tornado length. 

The appearance of the tornadoes in the AIA 171\,\AA\ images is  
affected by the spatial resolution of the instrument and the formation of 
the intensity in the passband. 
The tornadoes certainly have a fine-structure on smaller spatial scales, which 
is more clearly visible with other diagnostics. 
For instance, our off-limb H$\alpha$ observations in Fig.~\ref{fig:halphalimb}
reveal thin threads with widths of $\leq{}0.5$\arcsec.
Already \citet{2005SoPh..226..239L} and \citet{2012ApJ...745..152A} reported on 
SST observations of thread-like structures with widths of $\leq 0.3$\arcsec. 
%

\subsection{Connection to photospheric magnetic fields.} 
\label{sec:magnet}

A polarity inversion line (PIL, also referred to as `neutral line') 
separates regions on the Sun in which either one polarity 
dominates and is often co-located with a filament channel 
\citep[cf.][]{1998Natur.396..440Z}. 
We analysed the HMI magnetograms (see Fig.~\ref{fig:magnetogram}) 
and found that tornado groups are typically arranged along PILs in 
connection with a filament. 
About one third of the  tornadoes in our sample (see Sect.~\ref{sec:abudis}) 
are so close to the limb that a PIL cannot reliably be determined. 
However, there are some clear cases for which the PIL extends notably 
on the disk. 
We find that at least 86\,\% of all identified tornadoes are located close to a PIL. 
It cannot be ruled out that the remaining cases are close to a 
(less obvious) PIL that was not clearly detected here.

It is hard to connect the footpoints of 171~\AA~AIA tornadoes to counterparts in 
the HMI magnetogram. 
We find no pronounced magnetic field concentrations close to where the footpoints 
are expected but, given the limited spatial resolution of HMI, only rather 
small-scale magnetic field concentrations. 
This is in line with the finding by \citet{2005SoPh..226..239L} who show 
that the threads of quiescent filaments, 
whose footpoints  we identify as tornadoes (see Sect.~\ref{sec:disc}), 
are rooted in weak magnetic fields  
\citep[see also][and references therein]{2010SSRv..151..333M}. 
According to \citet{2005SoPh..227..283L}, the majority \mbox{($\sim2/3$)} 
of the footpoints is located within the boundaries of the magnetic network 
in the photosphere, while the rest is connecting to weak fields in  
internetwork regions {\citep[cf.][]{1973SoPh...29..399P}.}

Furthermore, we find groups of small magnetic flux concentrations with 
alternating polarity along the PIL. 
A possible explanation is that individual filament threads are rooted along 
the PIL and connect the different polarities over intermediate distances so 
that the loop tops compose an effectively longer filament spine.
This hypothesis has to be tested through high-resolution magnetic field 
measurements of the kind demonstrated by \citet{2013hsa7.conf..786O}  
for a quiescent hedgerow prominence.

\subsection{Eruption events}
\label{sec:erupt}

There are (at least) 36 tornadoes in our sample (18\,\%) that clearly end with an 
eruption.
These tornadoes are organized in groups that form the legs of 
filament spines although the filament is not always clearly visible, e.g. when a group 
only appears at the limb shortly before eruption. 
Some prominences may only become visible shortly before their eruption 
\citep{2001SoPh..202..293E}.
Eruptions are best seen for the examples at the limb so that the number detected 
here is most likely a lower limit only.

\begin{figure}[t!]
\begin{center}
\vspace*{-4mm}
\includegraphics[width=6.8cm]{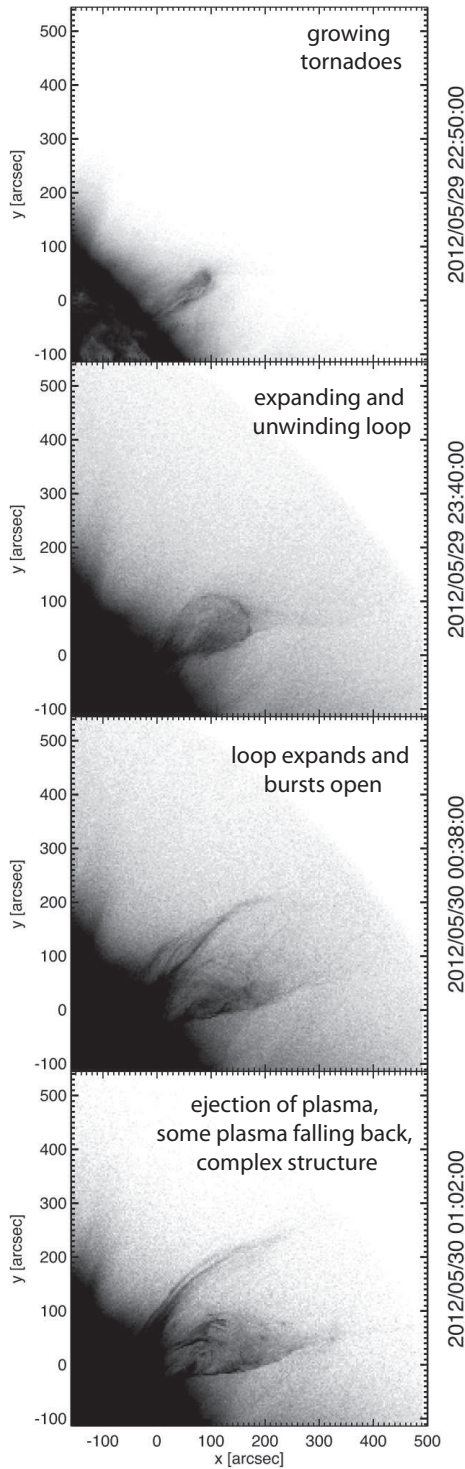}
\caption{Eruption of the tornado group that is shown in the upper row of 
Fig.~\ref{fig:tornado}. 
The uppermost panel shows the group shortly before 
eruption, whereas the other three panels show different stages of the eruption 
from top to bottom. 
The panels show inverted AIA 171\,\AA\ images, which have been enhanced for better 
visibility of the ejected plasma. 
The coordinates are centered on the footpoint of the initial tornado group.
\label{fig:eruption}
}
\end{center}
\end{figure}

As already noted by \citet{2012ApJ...756L..41S}, tornadoes erupt together with the 
filament, which is expected if they are indeed part of the filament. 
The  group shown in Fig.~\ref{fig:eruption} consists of 5 tornadoes which appear 
close together at the west limb. 
The connecting filament spine is visible in time sequences together 
with plasma that appears to spiral upwards through the tornadoes into the filament 
spine. 
The tornadoes at the limb seem to grow in height prior to eruption, resulting 
in an extended absorption feature in the AIA\,171\,\AA\ image (see 
uppermost panel in Fig.~\ref{fig:eruption}).

The eruption of the example in Fig.~\ref{fig:eruption} begins with a 
pronounced loop, which most likely connects two of the former tornadoes. 
The lower parts of the loop appear to intersect, which may explain the observation 
that the loop unwinds and expands afterwards. 
The loop top seems to rise with an apparent speed that increases from 10\,km/s to 
30\,km/s until it reaches a height of $\sim200$\,Mm beyond the limb. 
Then, the loop bursts open and plasma threads are ejected outwards with high speeds 
on the order of 200\,km/s. 
We also observe material falling back down afterwards. 
Shortly after the eruption of this loop more loops appear, which are rooted at 
slightly different locations on the surface. 
We interpret it such that individual parts of the filaments, which connect 
different tornadoes in individual loops,  erupt one after another. 
The eruption of the first loop possibly triggers the eruption of the other loops.

The initial growth of a tornado together with the rise of the filament and the 
cavity has also been observed  by \citet{2012ApJ...756L..41S}. 
A possible explanation is that the rotation of the magnetic field, which is most 
likely seen as a tornado, produces an increasing twist of the filament, which 
eventually becomes unstable and erupts 
\citep[cf., e.g.,][]{2013ApJ...764...91S,2013AJ....145..153Y}. 
Eruptions due to the injection of helicity are also discussed in detail by, e.g., 
\citet{2005ARA&A..43..103Z} and \citet{2006ApJ...644..575Z}. 
The existence of resulting helical structures during prominence eruptions has  
been reported repeatedly \citep[e.g.,][and references therein]{2012ApJ...756L..41S}.

\begin{figure*}[t!]
\centering
\includegraphics[width=\textwidth]{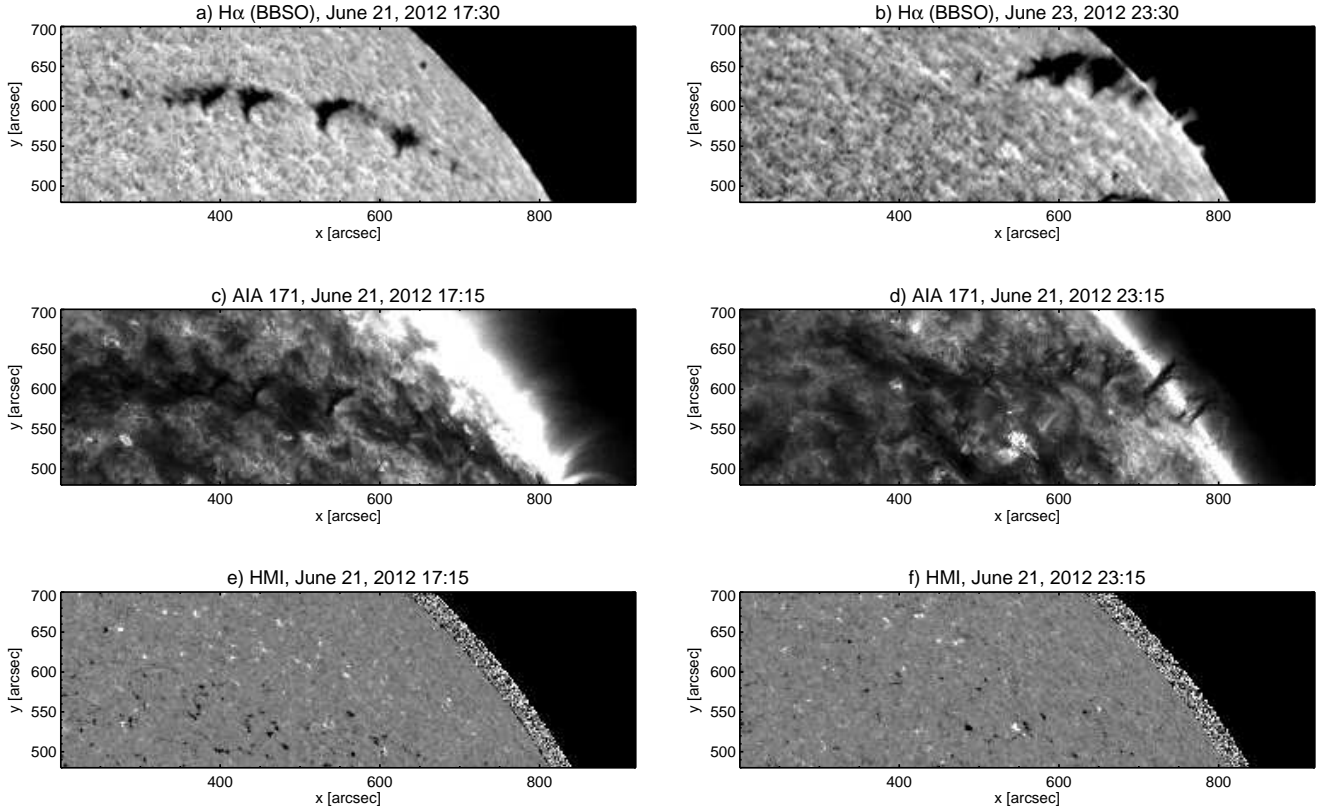}
\caption{The relation of tornadoes to filament channels and barbs illustrated for 
an exemplary tornado group at two different times (left column and right column). 
All panels show the same region on the Sun and display the BBSO H$\alpha$ images 
(top), the AIA~171\,\AA\ images (middle), and the corresponding HMI magnetograms (bottom). 
\label{fig:barbs}}
\end{figure*}

\section{Discussion}
\label{sec:disc}

\paragraph{Tornadoes, filaments and barbs.}
The conclusion that giant tornadoes are parts of filaments is supported by the 
fact that the vast majority of the tornadoes in our sample and possibly all of 
them are located along filament channels as seen in H$\alpha$ images  
and along polarity inversion lines (see Sect.~\ref{sec:magnet}),
which are indicators of filaments in the atmosphere above (see Fig.~\ref{fig:barbs}). 
Already the spatial distribution of the tornadoes gives a hint of their nature. 
They are distributed mostly in the mid-latitudes, probably close to sunspot 
latitude belts, or at their boundaries, and  in elongated groups. 
The fact that there are only few tornadoes observed close to 
the poles (see Fig.~\ref{fig:overview}) also indicates that they are not 
entirely a quiet Sun phenomena, so rather intermediary.
The more complex structure of active regions makes it more 
difficult to discern tornadoes there, which introduces a potential detection bias 
concerning events related to active region prominences.  
Most of the filaments with tornadoes reported here are therefore  
characterized as quiescent.

The example in Fig.~\ref{fig:halphalimb} illustrates how different the same 
event appears in different diagnostics. 
The AIA 171\,\AA\ images show tornadoes as narrow elongated dark features. 
The spatial resolution of the observation certainly affects the determined 
length of the features since the parts towards 
the narrow footpoint of a tornado might not be resolved.   
Further differences are caused by the different formation of the intensity 
captured by different diagnostics so that effectively different parts of 
a tornado are mapped. 
We argue here that H$\alpha$ line core images mostly show the lower parts of 
tornadoes.  
\citet{2012ApJ...752L..22L} and \citet{2012ApJ...756L..41S} identify their 
tornadoes with `barbs'. 
They describe barbs as the observational signatures of tornadoes when seen from the side
\citep[cf. middle row in Fig.~10 by][]{1998SoPh..182..107M}. 
For many events in our data set, we indeed find funnel-like dark features in 
H$\alpha$ at the exact same locations as the tornadoes (see the example in 
Fig.~\ref{fig:barbs}).
We would like to remark that these funnel-like features may appear differently compared to 
what is referred to as 'barbs' by other authors 
\citep[see, e.g.,][and references therein]{1999SoPh..185...87M,2013SoPh..tmp..146J}. 
The example in Fig.~\ref{fig:barbs} represents a situation, where the filament spine 
is barely visible in H$\alpha$, instead only a row of funnel-like features is seen. 
However, for other filaments with clearly visible spine it is obvious that giant 
tornadoes connect smoothly to the filament spine (see Fig.~\ref{fig:halphalimb}f) 
and that they seem to be composed of thin threads that extend from the chromosphere
(see Fig.~\ref{fig:halphalimb}c).
These  characteristics are also found for barbs. 
Furthermore, we note that the barbs studied in detail by \citet{2013SoPh..282..147L} 
in AIA~171\,\AA\ and H$\alpha$ images look identical to the tornadoes discussed here.
For the 58~barbs in their sample, they derive an average length of 23.7\,Mm, 
an average width of 2.1\,Mm, and an average lifetime of 17.2\,h. 
While the average length of barbs agrees well with the one for the 201~giant tornadoes derived here, 
barbs  appear to be somewhat thinner and short-lived compared to the giant tornadoes, 
although this might be due to differences in the measurement details
and/or the different sizes of the samples in the two studies.  
In conclusion, our results suggest that giant tornadoes are connected to 
funnel-like dark features close to filaments/prominences 
in H$\alpha$ images, which supports the view by \citet{2012ApJ...756L..41S}.   
However, it has still to be shown if these features are equivalent to barbs 
or if this only holds under certain conditions and/or for certain stages 
during the evolution of filaments. 
For instance, tornadoes might simply be a sub-group of rotating barbs. 
A comprehensive definition of the terms `barb' and `giant tornado' might be key for 
solving this apparent controversy. 
Whether `barb' or `tornado' is the more suitable name is directly linked to 
the question if these events are truly rotating or not.

As seen in section \ref{sec:offlimbtornado}, most of the SDO channels show 
tornadoes as dark absorption-like structures, indicating that it corresponds 
to plasma with temperatures outside the (coronal) ranges where the contribution 
functions of the channels are sensitive. 
Only the 304~\AA~AIA channel shows the tornado plasma in emission, and this 
indicates that part of it is at a temperature close to $8\times10^4~$K. 
The off-limb tornado observed with the SST indicates a very cool temperature 
component with an average upper-limit temperature of 7\,000~K. 
The threaded structure observed with the SST cannot be resolved with AIA, 
which prevents us from drawing conclusions about the thermal structure 
of tornadoes in general. 
The observed co-located emission in 304~\AA~AIA and H$\alpha$ images could be 
the result of a 'prominence corona transition region' \citep{2007A&A...469.1109P} 
or of a threaded sub-substructure with a mixture of cool and hot plasma.  
It is clear, however, that the very low temperatures agree with temperature 
diagnostics in prominences \citep{1985SoPh..100..415H}, which therefore strongly 
supports our view that giant tornadoes constitute sources and sinks to prominences. 
The existence of such flows can be expected if barbs and tornadoes would indeed 
be connected as suggested by \citet{2012ApJ...756L..41S}. 
The existence of upflows and/or downflows co-spatial to vortex motions has been 
reported repeatedly for magnetic structures of different sizes in the solar 
atmosphere \citep[e.g.,][]{1974IAUS...57..323B,1980HiA.....5..557B,1988Natur.335..238B,2008ApJ...687L.131B,2009A&A...507L...9W,2009Sci...323.1582J,2010A&A...510L...1K,2011ApJ...741L...7Z}. 
A final verification of this hypothesis requires further high-resolution 
observations of tornadoes and prominences.

In this respect, the finding that barbs are often already observed before the 
filament spine becomes visible 
\citep[e.g.][]{2005ASPC..346..219P,2012ApJ...752L..22L,2012ApJ...756L..41S} 
could be interpreted such that they serve as initial plasma sources
for prominences.
We also find a group of tornadoes co-located with the  forming prominence
analysed by \citet{2012ApJ...758L..37B}, although it is not clear  if these 
tornadoes play a role in the formation process.
Furthermore, the Doppler shifts that we determined in Sect.~\ref{sec:offlimbtornado} 
are of the same order as the speeds in counter-streaming flows in barbs as 
derived by \citet{1998Natur.396..440Z}.

The Interface Region Imaging Spectrograph (IRIS), 
which was launched in June 2013, is a promising instrument for detecting the lower to upper 
chromospheric parts of tornadoes and might shed light on many of the still open 
questions. 
IRIS is specially designed for spectrometric observations of the solar atmosphere 
in UV lines with high spatial and temporal resolutions. 
For instance, observations in the \mbox{\ion{Mg}{2}~h $\&$ k} lines 
($\log\mathrm{T}=4.0~\mathrm{K}$) will have a cadence of $1-2~$s, 
a spatial resolution of $0.4\arcsec$ (with a field of view of 
$0.3\arcsec\times40\arcsec$), an effective area of $0.25~\mathrm{cm}^{2}$ and a 
wavelength resolution of 80~m\AA.

\paragraph{The role of tornadoes in prominence eruptions.}
Prominence eruptions are usually analyzed numerically and analytically in a 
scenario in which prominences are composed by a helical flux rope anchored in the 
photosphere \citep{Kippenhahn_Schluter_1957ZA.....43...36K,1995ApJ...443..818L}. 
The magnetic twist found in this structure is usually attributed to gradual 
changes in the photospheric boundaries, compromising its stability by the 
increase of coronal magnetic stress. 
The latter can be achieved with sub-photospheric torsional Alfv\'en waves 
bringing twist to the coronal field or through reconnection between sheared 
magnetic loops from converging flows at the PIL, and occurs in a timescale of a 
few days \citep{2010ApJ...718.1388D,2012ApJ...760...31K}.
In this classical view, the prominence is generally seen as one whole structure. 
In the present work we have seen that a tornado can be considered as a magnetic 
entity having one end fixed at the photosphere and the other end connected to a 
large mass reservoir such as a prominence. 
In this view, tornadoes can not only be considered as legs to prominences but also 
as mass regulatory systems for the latter. 
Furthermore, the presented observations suggest a scenario in which tornadoes 
play an important role on the stability of the prominence. 
Along this view, it is important to consider whether conditions leading to a 
prominence eruption, like the injection of helicity  
\citep[cf., e.g.,][]{2005ARA&A..43..103Z,2006ApJ...644..575Z}, may crucially 
depend on tornadoes.

The presence of rotation in tornadoes reported in this work and in 
\citet{2012ApJ...756L..41S} provides an image of a tornado reminiscent of that 
of the prominence flux rope with the main difference being that its axis is vertical 
instead of horizontal. 
As a first approximation we can therefore consider a tornado as a vertical flux 
rope with a fixed lower end (on the photosphere) and a somewhat looser upper end 
with higher degree of freedom rooted to a large mass reservoir. 
Along this view a tornado can become unstable and erupt from the loss of 
equilibrium and ideal MHD instability 
\citep{1995ApJ...446..377F,2006PhRvL..96y5002K,2010ApJ...718.1388D}. 
A possible scenario in which such loss of equilibrium could happen is that set by 
the kink instability. 
If the winding of magnetic field lines along the tornado exceeds a threshold, the 
structure is deformed into a helical structure because of the current driven 
instability known as the kink instability 
\citep{1982QB539.M23P74...,1998ApJ...493L..43M}. 
\citet{1979SoPh...64..303H} and \citet{1996ApJ...469..954L} analyze the critical 
twist angle above which a flux tube is unstable against a kink mode. 
Values between $2\,\times\,\pi$ and $10\,\times\,\pi$ are found, depending on the 
magnetic field topology (uniformly twisted force-free field or other). 
For the case of the off-limb tornado observed by the SST, Doppler shifts 
around 20\,km\,s$^{-1}$ are found, implying an angular velocity of 0.004~radians 
per second for the observed radius of $\sim5$~Mm. 
In a low-beta plasma environment these speeds would imply a very fast magnetic field 
winding of one revolution every half an hour. 
It is therefore likely that a kink instability may be triggered. 
Along this view, we predict that the amount of twist found in a tornado is 
inversely proportional to its lifetime.

\begin{figure}
\centering
\includegraphics[width=8cm]{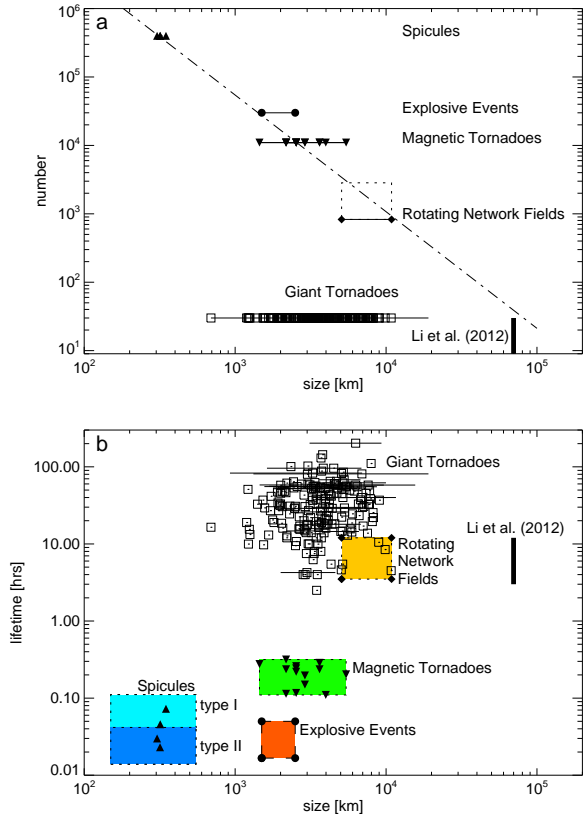}
\caption{Comparison of different rotating magnetic phenomena on the Sun: 
Giant tornadoes (squares), Rotating Network Fields (diamonds), magnetic tornadoes 
(triangles, tip downward), Explosive Events (circles), and spicules (triangles, 
tip upwards). 
\textbf{a)}~Estimated total numbers of events as a function of 
their characteristic spatial scale.
\textbf{b)}~Lifetimes versus characteristic size. 
Measurements of the same giant tornado at different times are connected by solid lines. 
The individual data points are explained in the text in Sect.~\ref{sec:disc}. 
The dot-dashed line in the upper panel represents a double-logarithmic 
power law for all phenomena except for the giant tornadoes.
\label{fig:power}}
\end{figure}

\paragraph{Rotating magnetic fields on different spatial scales.} 
\label{sec:powerlaw}
Solar tornadoes are essentially spatially confined rotating magnetic field 
structures that seem to exist on a large range of spatial scales. 
The giant tornadoes discussed here tend to be larger than the magnetic tornadoes 
presented by \citet{2012Natur.486..505W}, which have widths  
between 2\arcsec\ and 5.5\arcsec, although 57\,\% of all cases 
have effective widths in that range, too (see Fig.~\ref{fig:width}). 
The off-limb tornado discussed in Sect.~\ref{sec:offlimbtornado} has with 
a width of 6\arcsec\,-\,8\arcsec\ at its base and is thus only slightly larger than 
the so far largest observed chromospheric swirl with a width~of~5.5\arcsec.
These two phenomena might therefore  be related if not even just 
different sizes of the same scalable phenomenon. 
We investigate this hypothesis by plotting the abundance and the lifetimes
of giant and magnetic tornadoes over their characteristic size (see Fig.~\ref{fig:power}). 
The characteristic size is here defined as the typical diameter 
perpendicular to the rotation axis in the upper atmosphere. 
Based on the diameters derived in this study (see Sect.~\ref{sec:tornadosize} 
and Fig.~\ref{fig:width}c), giant tornadoes are larger and persist longer but 
appear in smaller numbers than  magnetic tornadoes. 
It should be noted that magnetic tornadoes have only been detected near disk-center 
as chromospheric swirls so far, while the AIA~171\,\AA\ imprint of giant tornadoes is 
difficult to see there, resulting in much reduced number of detections in the 
central region of the disk (see Fig.~\ref{fig:overview}).

\citet{2012ApJ...752L..22L} report on an extremely large tornado with a 
complicated fine-structure, which seems to be much larger than the examples 
discussed here. 
\citeauthor{2012ApJ...752L..22L} determine a radius of 35\,Mm for the circular 
trajectory of a plasma blob, which they track in the AIA\,171\,\AA\ 
images, and specify a duration of 3~hours for the clearest revolution. 
By revisiting these data, we find that the tornado described by 
\citeauthor{2012ApJ...752L..22L} can be followed for almost half a day and 
that it is co-located with a long-lived group of tornadoes, which are of the 
same type as those presented here (e.g., in Fig.~\ref{fig:tornado}). 
Among the giant tornadoes in the group, there is a pair, which is located not 
far from the apparent footpoint of the large-scale helical structure. 
This tornado pair seems to already exist at least 38~hours (or possibly days) 
before the helical event but it is no longer visible once the large-scale tornado 
develops. 
The connection between the giant tornadoes of the type described here and the 
large-scale helical event reported by \citeauthor{2012ApJ...752L..22L} is 
not clear yet. 
The latter seems to be connected to the photosphere by several threads whereas 
such a fine-structure may not be resolved for the giant tornadoes. 
We mark the corresponding characteristic size of 70\,Mm with a thick vertical 
line in Fig.~\ref{fig:power} because the abundance of this type of tornado is 
unclear yet. 
It might be as high as the abundance of the giant tornadoes presented in this work 
or possibly less. 

Tornadoes are not the only observed examples of rotating magnetic structures. 
Phenomena that combine rotation and magnetic fields are known to span a large range of 
spatial scales in the atmosphere of the Sun. 
Among them are, e.g., 
so-called `Explosive Events' (EEs) or swirling transition region jets 
\citep{2012SoPh..280..417C,2011A&A...532L...9C,1989ApJ...345L..95D,2004ESASP.547..215I}, 
EUV cyclones or `Rotating Network Magnetic Fields' \citep[RNFs,][]{2011ApJ...741L...7Z} 
and macrospicules  
\citep{1997SoPh..175..457P,2000A&A...355.1152B,2009ApJ...704.1385S,2011A&A...535A..58M}. 
More examples may or may not include the rotation and/or helicity in coronal jets 
\citep{2008ApJ...680L..73P,2009SoPh..259...87N,2011ApJ...735L..18L,2011ApJ...735L..43S}, 
motions in spicules \citep{2008ASPC..397...27S,2012ApJ...752L..12D,2013ApJ...769...44S}, 
macroscopic EUV jets \citep{2010ApJ...722.1644S}, 
`spinning magnetic twist jet(s)' \citep{1997ESASP.404..103S}
and rotating sunspots \citep{2012ApJ...761...60V}. 
In the following, we compare these phenomena to solar tornadoes.

For rotating magnetic network fields (or EUV cyclones), we adopt
the average lifetime of 3.5~hours given by \citet{2011ApJ...741L...7Z}, 
although they present two examples with lifetimes of 12 and 9~hours, respectively. 
The characteristic width is estimated to be in the range between 7\arcsec\ and 15\arcsec. 
\citeauthor{2011ApJ...741L...7Z} state that 5600~RNFs are present every day 
across the whole Sun. 
Based on the average lifetime of 3.5~hours, we conclude that about 830~RNFs 
would be present on the Sun at all times. 
This number would increase to 2800 for a lifetime of 12~hours 
(see dotted line for RNFs in Fig.~\ref{fig:power}a).

Explosive Events (EEs, also known as Swirling Transition Region Jets) 
are observed spectroscopically in radiation that originates  
from the transition region. 
\cite{2012SoPh..280..417C} 
characterize EEs as bi-directional flows, 
which last typically for one to three minutes, have spatial extents of 
1500\,km to 2500\,km and exhibit Doppler velocities of $\pm (50 - 150)\,\mathrm{km\,s}^{-1}$.  
According to \citet{2004A&A...427.1065T}, 30\,000~EEs exist at all times.

Spicules would be placed at the  small end of spatial scales but it is not clear 
yet if the torsional motions reported by \citet{2012ApJ...752L..12D} 
qualify spicules as rotating magnetic structures or if the detected motions are 
related to Alfv{\'e}n waves that propagate on a non-rotating magnetic field structure. 
For spicules, we adopt the  mean values for lifetimes and widths from 
\citet[][see their Tables~3 and~4]{2012ApJ...759...18P}. 
Spicules are the most abundant and shortest-lived events considered here.

At first glance, the total number $N_i$ of most of the different phenomena seems to 
be correlated to their characteristic size $\bar{x}_i$.
The correlation can be approximated with the double-logarithmic 
power law
\begin{equation} 
N_i = 7.1\,\times\,10^9 + \bar{x}_i ^{\ \ -1.7}\qquad, 
\end{equation}
which fits for the considered phenomena with exception of the giant tornadoes  
(see dot-dashed line in Fig.~\ref{fig:power}).
For some so far unknown reason, giant tornadoes seem not numerous enough and too 
narrow to fit the trend that is found for the other phenomena. 
Possible reasons might be connected to the finding that giant tornadoes are the 
legs of prominences, whereas the other phenomena are individual events that 
are not directly connected to large-scale structures.

\paragraph{Implications for other stars.} 
Tornadoes can be expected to exist also in the atmospheres of cooler stars because 
basic ingredients for a tornado, namely photospheric vortex flows and magnetic 
fields, are most likely common there, too. 
Photospheric vortex flows have been found in a numerical simulation of a 
M-type dwarf star by \citet{2006A&A...459..599L}. 
More recently, \citet{2013AN....334..137W} report on a first example of a 
(small-scale) magnetic tornado in a numerical radiation magnetohydrodynamics 
simulation that includes magnetic fields and a chromosphere for a M-type dwarf 
star. 
It remains to be seen if also giant tornadoes can form in the atmospheres 
of cool stars.

\section{Conclusions}
\label{sec:conc}

Our findings support the suggestion by \citet{2012ApJ...756L..41S} that giant 
tornadoes are connected to filaments. 
We find many examples of tornado groups, which first appear as coherent filaments 
on disk and only become visible as individual tornadoes when the structure rotates  
closer towards the solar limb (less than $\sim 160$\arcsec). 
The tornadoes appear to  connect smoothly to the filament spine in the atmosphere 
above, which has been observed for barbs before \citep[e.g][]{1998Natur.396..440Z}. 
Both the SDO and SST observations presented here suggest the existence of a cool  
plasma flow along the tornado axis with temperatures typical to the lower to upper 
chromosphere, although this finding has to be confirmed in a future more detailed 
study. 
This result strongly supports our view that tornadoes can act as sources and sinks 
for prominences.
The SST observations also suggest that tornadoes rotate around their vertical 
axis. 
The numerical model of small-scale solar tornadoes by \citet{2012Natur.486..505W}, 
in which tornadoes are driven by photospheric vortex flows, exhibits flows that 
spiral both upwards and downwards with speeds comparable to the values found for 
filament barbs. 
In agreement with \citet{2012ApJ...756L..41S}, 
we conclude that this rotation could lead to an increasing twist of the magnetic 
structure of the overlying prominence until it becomes unstable and erupts. 
We observed several examples of tornado groups, which are unambiguously  
connected to erupting prominences. 
It seems therefore likely that the rotation of tornadoes could be a common trigger 
of prominence eruptions, although this conclusion has to be investigated further. 
Erupting prominences cause Coronal Mass Ejections (CMEs) which often 
eject large amounts of coronal plasma into the interplanetary space
\citep{1994SSRv...70..215B}. 
Monitoring of giant tornadoes might therefore provide a way to forecast CMEs, in 
case their role as trigger of eruptions is confirmed by future studies. 

\acknowledgments
The authors thank O.~Engvold for helpful discussions. 
SW and AB thank the Faculty of Mathematics and Natural Sciences of the University 
of Oslo and IAESTE for support. 
SW acknowledges support from the Research Council of Norway 
(grants No.\,208011/F50 and 221767/F20). 
This research has made use of the Solar Dynamics Observatory, the Big Bear Solar 
Telescope, ESA's JHelioviewer, NASA's Astrophysics Data System, and the Virtual 
Solar Observatory.  
The Swedish 1-m Solar Telescope is operated on the island of La Palma by the 
Institute for Solar Physics of the Royal Swedish Academy of Sciences in the Spanish 
Observatorio del Roque de los Muchachos of the Instituto de Astrof{\'i}sica de Canarias.

{\it Facilities:} \facility{SST (CRISP)}, \facility{SDO (AIA, HMI)}, \facility{BBSO}


\begin{thebibliography}{86}
\expandafter\ifx\csname natexlab\endcsname\relax\def\natexlab#1{#1}\fi

\bibitem[{{Antolin} \& {Rouppe van der Voort}(2012)}]{2012ApJ...745..152A}
{Antolin}, P. \& {Rouppe van der Voort}, L. 2012, \apj, 745, 152

\bibitem[{{Anzer} \& {Heinzel}(2005)}]{2005ApJ...622..714A}
{Anzer}, U. \& {Heinzel}, P. 2005, \apj, 622, 714

\bibitem[{{Banerjee} {et~al.}(2000){Banerjee}, {O'Shea}, \&
  {Doyle}}]{2000A&A...355.1152B}
{Banerjee}, D., {O'Shea}, E., \& {Doyle}, J.~G. 2000, \aap, 355, 1152

\bibitem[{{Berger} {et~al.}(2012){Berger}, {Liu}, \&
  {Low}}]{2012ApJ...758L..37B}
{Berger}, T.~E., {Liu}, W., \& {Low}, B.~C. 2012, \apjl, 758, L37

\bibitem[{{Bonet} {et~al.}(2008){Bonet}, {M{\'a}rquez}, {S{\'a}nchez Almeida},
  {Cabello}, \& {Domingo}}]{2008ApJ...687L.131B}
{Bonet}, J.~A., {M{\'a}rquez}, I., {S{\'a}nchez Almeida}, J., {Cabello}, I., \&
  {Domingo}, V. 2008, \apjl, 687, L131

\bibitem[{{Bothmer} \& {Schwenn}(1994)}]{1994SSRv...70..215B}
{Bothmer}, V. \& {Schwenn}, R. 1994, \ssr, 70, 215

\bibitem[{{Brandt} {et~al.}(1988){Brandt}, {Scharmer}, {Ferguson}, {Shine}, \&
  {Tarbell}}]{1988Natur.335..238B}
{Brandt}, P.~N., {Scharmer}, G.~B., {Ferguson}, S., {Shine}, R.~A., \&
  {Tarbell}, T.~D. 1988, \nat, 335, 238

\bibitem[{{Brueckner}(1980)}]{1980HiA.....5..557B}
{Brueckner}, G.~E. 1980, Highlights of Astronomy, 5, 557

\bibitem[{{Bruzek}(1974)}]{1974IAUS...57..323B}
{Bruzek}, A. 1974, in IAU Symposium, Vol.~57, Coronal Disturbances, ed. G.~A.
  {Newkirk}, 323

\bibitem[{{Bruzek} \& {Durrant}(1977)}]{1977ASSL...69.....B}
{Bruzek}, A. \& {Durrant}, C.~J., eds. 1977, Astrophysics and Space Science
  Library, Vol.~69, {Illustrated glossary for solar and solar-terrestrial
  physics}

\bibitem[{{Curdt} \& {Tian}(2011)}]{2011A&A...532L...9C}
{Curdt}, W. \& {Tian}, H. 2011, \aap, 532, L9+

\bibitem[{{Curdt} {et~al.}(2012){Curdt}, {Tian}, \&
  {Kamio}}]{2012SoPh..280..417C}
{Curdt}, W., {Tian}, H., \& {Kamio}, S. 2012, \solphys, 280, 417

\bibitem[{{De Pontieu} {et~al.}(2012){De Pontieu}, {Carlsson}, {Rouppe van der
  Voort}, {Rutten}, {Hansteen}, \& {Watanabe}}]{2012ApJ...752L..12D}
{De Pontieu}, B., {Carlsson}, M., {Rouppe van der Voort}, L.~H.~M., {et~al.}
  2012, \apjl, 752, L12

\bibitem[{{D{\'e}moulin} \& {Aulanier}(2010)}]{2010ApJ...718.1388D}
{D{\'e}moulin}, P. \& {Aulanier}, G. 2010, \apj, 718, 1388

\bibitem[{{Dere} {et~al.}(1989){Dere}, {Bartoe}, {Brueckner}, \&
  {Recely}}]{1989ApJ...345L..95D}
{Dere}, K.~P., {Bartoe}, J.-D.~F., {Brueckner}, G.~E., \& {Recely}, F. 1989,
  \apjl, 345, L95

\bibitem[{{Domingo} {et~al.}(1995){Domingo}, {Fleck}, \&
  {Poland}}]{1995SoPh..162....1D}
{Domingo}, V., {Fleck}, B., \& {Poland}, A.~I. 1995, \solphys, 162, 1

\bibitem[{{Engvold}(1998)}]{1998ASPC..150...23E}
{Engvold}, O. 1998, in Astronomical Society of the Pacific Conference Series,
  Vol. 150, IAU Colloq. 167: New Perspectives on Solar Prominences, ed. D.~F.
  {Webb}, B.~{Schmieder}, \& D.~M. {Rust}, 23

\bibitem[{{Engvold} {et~al.}(2001){Engvold}, {Jakobsson}, {Tandberg-Hanssen},
  {Gurman}, \& {Moses}}]{2001SoPh..202..293E}
{Engvold}, O., {Jakobsson}, H., {Tandberg-Hanssen}, E., {Gurman}, J.~B., \&
  {Moses}, D. 2001, \solphys, 202, 293

\bibitem[{{Forbes} \& {Priest}(1995)}]{1995ApJ...446..377F}
{Forbes}, T.~G. \& {Priest}, E.~R. 1995, \apj, 446, 377

\bibitem[{{Harrison} {et~al.}(1995){Harrison}, {Sawyer}, {Carter}, {Cruise},
  {Cutler}, {Fludra}, {Hayes}, {Kent}, {Lang}, {Parker}, {Payne}, {Pike},
  {Peskett}, {Richards}, {Gulhane}, {Norman}, {Breeveld}, {Breeveld}, {Al
  Janabi}, {McCalden}, {Parkinson}, {Self}, {Thomas}, {Poland}, {Thomas},
  {Thompson}, {Kjeldseth-Moe}, {Brekke}, {Karud}, {Maltby}, {Aschenbach},
  {Br{\"a}uninger}, {K{\"u}hne}, {Hollandt}, {Siegmund}, {Huber}, {Gabriel},
  {Mason}, \& {Bromage}}]{1995SoPh..162..233H}
{Harrison}, R.~A., {Sawyer}, E.~C., {Carter}, M.~K., {et~al.} 1995, \solphys,
  162, 233

\bibitem[{{Hill} {et~al.}(2004){Hill}, {Bogart}, {Davey}, {Dimitoglou},
  {Gurman}, {Hourcle}, {Martens}, {Suarez-Sola}, {Tian}, {Wampler}, \&
  {Yoshimura}}]{2004SPIE.5493..163H}
{Hill}, F., {Bogart}, R.~S., {Davey}, A., {et~al.} 2004, in Society of
  Photo-Optical Instrumentation Engineers (SPIE) Conference Series, Vol. 5493,
  Society of Photo-Optical Instrumentation Engineers (SPIE) Conference Series,
  ed. P.~J. {Quinn} \& A.~{Bridger}, 163--169

\bibitem[{{Hirayama}(1985)}]{1985SoPh..100..415H}
{Hirayama}, T. 1985, \solphys, 100, 415

\bibitem[{{Hood} \& {Priest}(1979)}]{1979SoPh...64..303H}
{Hood}, A.~W. \& {Priest}, E.~R. 1979, \solphys, 64, 303

\bibitem[{{Innes}(2004)}]{2004ESASP.547..215I}
{Innes}, D.~E. 2004, in ESA Special Publication, Vol. 547, SOHO 13 Waves,
  Oscillations and Small-Scale Transients Events in the Solar Atmosphere: Joint
  View from SOHO and TRACE, ed. H.~{Lacoste}, 215

\bibitem[{{Jess} {et~al.}(2009){Jess}, {Mathioudakis}, {Erd{\'e}lyi},
  {Crockett}, {Keenan}, \& {Christian}}]{2009Sci...323.1582J}
{Jess}, D.~B., {Mathioudakis}, M., {Erd{\'e}lyi}, R., {et~al.} 2009, Science,
  323, 1582

\bibitem[{{Joshi} {et~al.}(2013){Joshi}, {Srivastava}, {Mathew}, \&
  {Martin}}]{2013SoPh..tmp..146J}
{Joshi}, A.~D., {Srivastava}, N., {Mathew}, S.~K., \& {Martin}, S.~F. 2013,
  \solphys

\bibitem[{{Kamio} {et~al.}(2010){Kamio}, {Curdt}, {Teriaca}, {Inhester}, \&
  {Solanki}}]{2010A&A...510L...1K}
{Kamio}, S., {Curdt}, W., {Teriaca}, L., {Inhester}, B., \& {Solanki}, S.~K.
  2010, \aap, 510, L1

\bibitem[{{Kippenhahn} \&
  {Schl{\"u}ter}(1957)}]{Kippenhahn_Schluter_1957ZA.....43...36K}
{Kippenhahn}, R. \& {Schl{\"u}ter}, A. 1957, \zap, 43, 36

\bibitem[{{Kliem} \& {T{\"o}r{\"o}k}(2006)}]{2006PhRvL..96y5002K}
{Kliem}, B. \& {T{\"o}r{\"o}k}, T. 2006, Physical Review Letters, 96, 255002

\bibitem[{{Kusano} {et~al.}(2012){Kusano}, {Bamba}, {Yamamoto}, {Iida},
  {Toriumi}, \& {Asai}}]{2012ApJ...760...31K}
{Kusano}, K., {Bamba}, Y., {Yamamoto}, T.~T., {et~al.} 2012, \apj, 760, 31

\bibitem[{{Lemen} {et~al.}(2012){Lemen}, {Title}, {Akin}, {Boerner}, {Chou},
  {Drake}, {Duncan}, {Edwards}, {Friedlaender}, {Heyman}, {Hurlburt}, {Katz},
  {Kushner}, {Levay}, {Lindgren}, {Mathur}, {McFeaters}, {Mitchell}, {Rehse},
  {Schrijver}, {Springer}, {Stern}, {Tarbell}, {Wuelser}, {Wolfson}, {Yanari},
  {Bookbinder}, {Cheimets}, {Caldwell}, {Deluca}, {Gates}, {Golub}, {Park},
  {Podgorski}, {Bush}, {Scherrer}, {Gummin}, {Smith}, {Auker}, {Jerram},
  {Pool}, {Soufli}, {Windt}, {Beardsley}, {Clapp}, {Lang}, \&
  {Waltham}}]{2012SoPh..275...17L}
{Lemen}, J.~R., {Title}, A.~M., {Akin}, D.~J., {et~al.} 2012, \solphys, 275, 17

\bibitem[{{Li} \& {Zhang}(2013)}]{2013SoPh..282..147L}
{Li}, L. \& {Zhang}, J. 2013, \solphys, 282, 147

\bibitem[{{Li} {et~al.}(2012){Li}, {Morgan}, {Leonard}, \&
  {Jeska}}]{2012ApJ...752L..22L}
{Li}, X., {Morgan}, H., {Leonard}, D., \& {Jeska}, L. 2012, \apjl, 752, L22

\bibitem[{{Lin} {et~al.}(2005{\natexlab{a}}){Lin}, {Engvold}, {Rouppe van der
  Voort}, {Wiik}, \& {Berger}}]{2005SoPh..226..239L}
{Lin}, Y., {Engvold}, O., {Rouppe van der Voort}, L., {Wiik}, J.~E., \&
  {Berger}, T.~E. 2005{\natexlab{a}}, \solphys, 226, 239

\bibitem[{{Lin} {et~al.}(2005{\natexlab{b}}){Lin}, {Wiik}, {Engvold}, {Rouppe
  van der Voort}, \& {Frank}}]{2005SoPh..227..283L}
{Lin}, Y., {Wiik}, J.~E., {Engvold}, O., {Rouppe van der Voort}, L., \&
  {Frank}, Z.~A. 2005{\natexlab{b}}, \solphys, 227, 283

\bibitem[{{Linton} {et~al.}(1996){Linton}, {Longcope}, \&
  {Fisher}}]{1996ApJ...469..954L}
{Linton}, M.~G., {Longcope}, D.~W., \& {Fisher}, G.~H. 1996, \apj, 469, 954

\bibitem[{{Liu} {et~al.}(2011){Liu}, {Deng}, {Liu}, {Ugarte-Urra}, {Wang}, \&
  {Wang}}]{2011ApJ...735L..18L}
{Liu}, C., {Deng}, N., {Liu}, R., {et~al.} 2011, \apjl, 735, L18

\bibitem[{{Low} \& {Hundhausen}(1995)}]{1995ApJ...443..818L}
{Low}, B.~C. \& {Hundhausen}, J.~R. 1995, \apj, 443, 818

\bibitem[{{Ludwig} {et~al.}(2006){Ludwig}, {Allard}, \&
  {Hauschildt}}]{2006A&A...459..599L}
{Ludwig}, H.-G., {Allard}, F., \& {Hauschildt}, P.~H. 2006, \aap, 459, 599

\bibitem[{{Mackay} {et~al.}(2010){Mackay}, {Karpen}, {Ballester}, {Schmieder},
  \& {Aulanier}}]{2010SSRv..151..333M}
{Mackay}, D.~H., {Karpen}, J.~T., {Ballester}, J.~L., {Schmieder}, B., \&
  {Aulanier}, G. 2010, \ssr, 151, 333

\bibitem[{{Mackay} {et~al.}(1999){Mackay}, {Longbottom}, \&
  {Priest}}]{1999SoPh..185...87M}
{Mackay}, D.~H., {Longbottom}, A.~W., \& {Priest}, E.~R. 1999, \solphys, 185,
  87

\bibitem[{{Martin}(1998)}]{1998SoPh..182..107M}
{Martin}, S.~F. 1998, \solphys, 182, 107

\bibitem[{{Matsumoto} {et~al.}(1998){Matsumoto}, {Tajima}, {Chou}, {Okubo}, \&
  {Shibata}}]{1998ApJ...493L..43M}
{Matsumoto}, R., {Tajima}, T., {Chou}, W., {Okubo}, A., \& {Shibata}, K. 1998,
  \apjl, 493, L43

\bibitem[{{Murawski} {et~al.}(2011){Murawski}, {Srivastava}, \&
  {Zaqarashvili}}]{2011A&A...535A..58M}
{Murawski}, K., {Srivastava}, A.~K., \& {Zaqarashvili}, T.~V. 2011, \aap, 535,
  A58

\bibitem[{{Nistic{\`o}} {et~al.}(2009){Nistic{\`o}}, {Bothmer}, {Patsourakos},
  \& {Zimbardo}}]{2009SoPh..259...87N}
{Nistic{\`o}}, G., {Bothmer}, V., {Patsourakos}, S., \& {Zimbardo}, G. 2009,
  \solphys, 259, 87

\bibitem[{{Orozco Su{\'a}rez} {et~al.}(2012){Orozco Su{\'a}rez}, {Asensio
  Ramos}, \& {Trujillo Bueno}}]{2012ApJ...761L..25O}
{Orozco Su{\'a}rez}, D., {Asensio Ramos}, A., \& {Trujillo Bueno}, J. 2012,
  \apjl, 761, L25

\bibitem[{{Orozco Su{\'a}rez} {et~al.}(2013){Orozco Su{\'a}rez}, {Asensio
  Ramos}, \& {Trujillo Bueno}}]{2013hsa7.conf..786O}
{Orozco Su{\'a}rez}, D., {Asensio Ramos}, A., \& {Trujillo Bueno}, J. 2013, in
  Highlights of Spanish Astrophysics VII, 786--791

\bibitem[{{Panasenco} {et~al.}(2013){Panasenco}, {Martin}, \&
  {Velli}}]{2013arXiv1307.2303P}
{Panasenco}, O., {Martin}, S.~F., \& {Velli}, M. 2013, ArXiv e-prints

\bibitem[{{Panesar} {et~al.}(2013){Panesar}, {Innes}, {Tiwari}, \&
  {Low}}]{2013A&A...549A.105P}
{Panesar}, N.~K., {Innes}, D.~E., {Tiwari}, S.~K., \& {Low}, B.~C. 2013, \aap,
  549, A105

\bibitem[{{Parenti} \& {Vial}(2007)}]{2007A&A...469.1109P}
{Parenti}, S. \& {Vial}, J.-C. 2007, \aap, 469, 1109

\bibitem[{{Patsourakos} {et~al.}(2008){Patsourakos}, {Pariat}, {Vourlidas},
  {Antiochos}, \& {Wuelser}}]{2008ApJ...680L..73P}
{Patsourakos}, S., {Pariat}, E., {Vourlidas}, A., {Antiochos}, S.~K., \&
  {Wuelser}, J.~P. 2008, \apjl, 680, L73

\bibitem[{{Pereira} {et~al.}(2012){Pereira}, {De Pontieu}, \&
  {Carlsson}}]{2012ApJ...759...18P}
{Pereira}, T.~M.~D., {De Pontieu}, B., \& {Carlsson}, M. 2012, \apj, 759, 18

\bibitem[{{Pettit}(1932)}]{1932ApJ....76....9P}
{Pettit}, E. 1932, \apj, 76, 9

\bibitem[{{Pettit}(1950)}]{1950PASP...62..144P}
{Pettit}, E. 1950, \pasp, 62, 144

\bibitem[{{Pevtsov} \& {Neidig}(2005)}]{2005ASPC..346..219P}
{Pevtsov}, A.~A. \& {Neidig}, D. 2005, in Astronomical Society of the Pacific
  Conference Series, Vol. 346, Large-scale Structures and their Role in Solar
  Activity, ed. K.~{Sankarasubramanian}, M.~{Penn}, \& A.~{Pevtsov}, 219

\bibitem[{{Pike} \& {Harrison}(1997)}]{1997SoPh..175..457P}
{Pike}, C.~D. \& {Harrison}, R.~A. 1997, \solphys, 175, 457

\bibitem[{{Pike} \& {Mason}(1998)}]{1998SoPh..182..333P}
{Pike}, C.~D. \& {Mason}, H.~E. 1998, \solphys, 182, 333

\bibitem[{{P{\l}ocieniak} \& {Rompolt}(1973)}]{1973SoPh...29..399P}
{P{\l}ocieniak}, S. \& {Rompolt}, B. 1973, \solphys, 29, 399

\bibitem[{{Priest}(1982)}]{1982QB539.M23P74...}
{Priest}, E.~R. 1982, {Solar magneto-hydrodynamics}, 74P

\bibitem[{{Scharmer} {et~al.}(2003{\natexlab{a}}){Scharmer}, {Bjelksjo},
  {Korhonen}, {Lindberg}, \& {Petterson}}]{2003SPIE.4853..341S}
{Scharmer}, G.~B., {Bjelksjo}, K., {Korhonen}, T.~K., {Lindberg}, B., \&
  {Petterson}, B. 2003{\natexlab{a}}, in Society of Photo-Optical
  Instrumentation Engineers (SPIE) Conference Series, ed. S.~L. {Keil} \& S.~V.
  {Avakyan}, Vol. 4853, 341

\bibitem[{{Scharmer} {et~al.}(2003{\natexlab{b}}){Scharmer}, {Dettori},
  {Lofdahl}, \& {Shand}}]{2003SPIE.4853..370S}
{Scharmer}, G.~B., {Dettori}, P.~M., {Lofdahl}, M.~G., \& {Shand}, M.
  2003{\natexlab{b}}, in Society of Photo-Optical Instrumentation Engineers
  (SPIE) Conference Series, Vol. 4853, Society of Photo-Optical Instrumentation
  Engineers (SPIE) Conference Series, ed. {S.~L.~Keil \& S.~V.~Avakyan},
  370--380

\bibitem[{{Scharmer} {et~al.}(2008){Scharmer}, {Narayan}, {Hillberg}, {de la
  Cruz Rodriguez}, {L{\"o}fdahl}, {Kiselman}, {S{\"u}tterlin}, {van Noort}, \&
  {Lagg}}]{2008ApJ...689L..69S}
{Scharmer}, G.~B., {Narayan}, G., {Hillberg}, T., {et~al.} 2008, \apjl, 689,
  L69

\bibitem[{{Scherrer} {et~al.}(2012){Scherrer}, {Schou}, {Bush}, {Kosovichev},
  {Bogart}, {Hoeksema}, {Liu}, {Duvall}, {Zhao}, {Title}, {Schrijver},
  {Tarbell}, \& {Tomczyk}}]{2012SoPh..275..207S}
{Scherrer}, P.~H., {Schou}, J., {Bush}, R.~I., {et~al.} 2012, \solphys, 275,
  207

\bibitem[{{Scullion} {et~al.}(2009){Scullion}, {Popescu}, {Banerjee}, {Doyle},
  \& {Erd{\'e}lyi}}]{2009ApJ...704.1385S}
{Scullion}, E., {Popescu}, M.~D., {Banerjee}, D., {Doyle}, J.~G., \&
  {Erd{\'e}lyi}, R. 2009, \apj, 704, 1385

\bibitem[{{Sekse} {et~al.}(2012){Sekse}, {Rouppe van der Voort}, \& {De
  Pontieu}}]{2012ApJ...752..108S}
{Sekse}, D.~H., {Rouppe van der Voort}, L., \& {De Pontieu}, B. 2012, \apj,
  752, 108

\bibitem[{{Sekse} {et~al.}(2013){Sekse}, {Rouppe van der Voort}, {De Pontieu},
  \& {Scullion}}]{2013ApJ...769...44S}
{Sekse}, D.~H., {Rouppe van der Voort}, L., {De Pontieu}, B., \& {Scullion}, E.
  2013, \apj, 769, 44

\bibitem[{{Shen} {et~al.}(2011){Shen}, {Liu}, {Su}, \&
  {Ibrahim}}]{2011ApJ...735L..43S}
{Shen}, Y., {Liu}, Y., {Su}, J., \& {Ibrahim}, A. 2011, \apjl, 735, L43

\bibitem[{{Shibata}(1997)}]{1997ESASP.404..103S}
{Shibata}, K. 1997, in ESA Special Publication, Vol. 404, Fifth SOHO Workshop:
  The Corona and Solar Wind Near Minimum Activity, ed. A.~{Wilson}, 103

\bibitem[{{Sterling} {et~al.}(2010){Sterling}, {Harra}, \&
  {Moore}}]{2010ApJ...722.1644S}
{Sterling}, A.~C., {Harra}, L.~K., \& {Moore}, R.~L. 2010, \apj, 722, 1644

\bibitem[{{Su} \& {van Ballegooijen}(2013)}]{2013ApJ...764...91S}
{Su}, Y. \& {van Ballegooijen}, A. 2013, \apj, 764, 91

\bibitem[{{Su} {et~al.}(2012){Su}, {Wang}, {Veronig}, {Temmer}, \&
  {Gan}}]{2012ApJ...756L..41S}
{Su}, Y., {Wang}, T., {Veronig}, A., {Temmer}, M., \& {Gan}, W. 2012, \apjl,
  756, L41

\bibitem[{{Suematsu} {et~al.}(2008){Suematsu}, {Ichimoto}, {Katsukawa},
  {Shimizu}, {Okamoto}, {Tsuneta}, {Tarbell}, \& {Shine}}]{2008ASPC..397...27S}
{Suematsu}, Y., {Ichimoto}, K., {Katsukawa}, Y., {et~al.} 2008, in Astronomical
  Society of the Pacific Conference Series, Vol. 397, First Results From
  Hinode, ed. S.~A. {Matthews}, J.~M. {Davis}, \& L.~K. {Harra}, 27

\bibitem[{{Teriaca} {et~al.}(2004){Teriaca}, {Banerjee}, {Falchi}, {Doyle}, \&
  {Madjarska}}]{2004A&A...427.1065T}
{Teriaca}, L., {Banerjee}, D., {Falchi}, A., {Doyle}, J.~G., \& {Madjarska},
  M.~S. 2004, \aap, 427, 1065

\bibitem[{{van Noort} {et~al.}(2005){van Noort}, {Rouppe van der Voort}, \&
  {L{\"o}fdahl}}]{2005SoPh..228..191V}
{van Noort}, M., {Rouppe van der Voort}, L., \& {L{\"o}fdahl}, M.~G. 2005,
  \solphys, 228, 191

\bibitem[{{Vemareddy} {et~al.}(2012){Vemareddy}, {Ambastha}, \&
  {Maurya}}]{2012ApJ...761...60V}
{Vemareddy}, P., {Ambastha}, A., \& {Maurya}, R.~A. 2012, \apj, 761, 60

\bibitem[{{Vissers} \& {Rouppe van der Voort}(2012)}]{2012ApJ...750...22V}
{Vissers}, G. \& {Rouppe van der Voort}, L. 2012, \apj, 750, 22

\bibitem[{{Wedemeyer} {et~al.}(2013){Wedemeyer}, {Ludwig}, \&
  {Steiner}}]{2013AN....334..137W}
{Wedemeyer}, S., {Ludwig}, H.-G., \& {Steiner}, O. 2013, Astronomische
  Nachrichten, 334, 137

\bibitem[{{Wedemeyer-B{\"o}hm} \& {Rouppe van der
  Voort}(2009)}]{2009A&A...507L...9W}
{Wedemeyer-B{\"o}hm}, S. \& {Rouppe van der Voort}, L. 2009, \aap, 507, L9

\bibitem[{{Wedemeyer-B{\"o}hm} {et~al.}(2012){Wedemeyer-B{\"o}hm}, {Scullion},
  {Steiner}, {van der Voort}, {de La Cruz Rodriguez}, {Fedun}, \&
  {Erd{\'e}lyi}}]{2012Natur.486..505W}
{Wedemeyer-B{\"o}hm}, S., {Scullion}, E., {Steiner}, O., {et~al.} 2012, \nat,
  486, 505

\bibitem[{{Yan} {et~al.}(2013){Yan}, {Pan}, {Liu}, {Qu}, {Xue}, {Deng}, {Ma},
  \& {Kong}}]{2013AJ....145..153Y}
{Yan}, X.~L., {Pan}, G.~M., {Liu}, J.~H., {et~al.} 2013, \aj, 145, 153

\bibitem[{{Zhang} \& {Liu}(2011)}]{2011ApJ...741L...7Z}
{Zhang}, J. \& {Liu}, Y. 2011, \apjl, 741, L7+

\bibitem[{{Zhang} {et~al.}(2006){Zhang}, {Flyer}, \&
  {Low}}]{2006ApJ...644..575Z}
{Zhang}, M., {Flyer}, N., \& {Low}, B.~C. 2006, \apj, 644, 575

\bibitem[{{Zhang} \& {Low}(2005)}]{2005ARA&A..43..103Z}
{Zhang}, M. \& {Low}, B.~C. 2005, \araa, 43, 103

\bibitem[{{Zirker}(1989)}]{1989SoPh..119..341Z}
{Zirker}, J.~B. 1989, \solphys, 119, 341

\bibitem[{{Zirker} {et~al.}(1998){Zirker}, {Engvold}, \&
  {Martin}}]{1998Natur.396..440Z}
{Zirker}, J.~B., {Engvold}, O., \& {Martin}, S.~F. 1998, \nat, 396, 440

\bibitem[{{Z{\"o}llner}(1869)}]{1869AN.....74..269Z}
{Z{\"o}llner}, F. 1869, Astronomische Nachrichten, 74, 269

\end{thebibliography}
\bibliographystyle{aa}

\end{document}